\documentclass[apj]{emulateapj}

\usepackage{textcomp}
\usepackage{amsmath}
\usepackage{graphicx}
\usepackage{natbib}

\def\MgII{Mg\,{\sc ii}}
\def\nar{New Astron. Rev.}

\shorttitle{$M_{\rm{BH}}-M_{\ast}$ Evolution and AGN Duty Cycle}
\shortauthors{Sun et al.}

\begin{document}
\slugcomment{\bf Draft: \today}

\title{Evolution in the Black Hole - Galaxy Scaling Relations 
and the Duty Cycle of Nuclear Activity in Star-Forming Galaxies}

\author{Mouyuan Sun\altaffilmark{1,2}}
\author{Jonathan R.\ Trump\altaffilmark{1,$\dagger$}}
\author{W.\ N.\ Brandt\altaffilmark{1}}
\author{B.\ Luo\altaffilmark{1}}
\author{David\ M.\ Alexander\altaffilmark{3}}
\author{Knud Jahnke\altaffilmark{4}}
\author{D.\ J.\ Rosario\altaffilmark{5}}
\author{Sharon\ X.\ Wang\altaffilmark{1}}
\author{Y. Q. Xue\altaffilmark{6}}

\altaffiltext{1}{Department of Astronomy \& Astrophysics and Institute for 
  Gravitation and the Cosmos, 525 Davey Lab, The Pennsylvania State University, 
  University Park, PA 16802, USA}
\altaffiltext{2}{Department of Astronomy and Institute of Theoretical 
  Physics and Astrophysics, Xiamen University, Xiamen, Fujian 361005, China}
\altaffiltext{3}{Institute for Computational Cosmology, Durham
  University, South Road, Durham, DH1 3LE, UK}
\altaffiltext{4}{Max Planck Institute for Astronomy, K\"onigstuhl 17,
  D-69117 Heidelberg, Germany}
\altaffiltext{5}{Max-Planck-Institut f\"{u}r extraterrestrische Physik (MPE),
  Giessenbachstrasse 1, D-85748 Garching bei M\"{u}nchen, Germany}
\altaffiltext{6}{Key Laboratory for Research in Galaxies and Cosmology, Center 
  for Astrophysics, Department of Astronomy, University of Science and Technology 
  of China, Chinese Academy of Sciences, Hefei, Anhui 230026, China}
\altaffiltext{$\dagger$}{Hubble Fellow}

%===================================================

\begin{abstract}
We measure the location and evolutionary vectors of 69 
\textit{Herschel}-detected broad-line 
active galactic nuclei (BLAGNs) in the $M_{\rm{BH}}-M_{\ast}$ 
plane. BLAGNs are selected from the COSMOS and CDF-S fields, and 
span the redshift range $0.2\leq z<2.1$. Black-hole masses are 
calculated using archival spectroscopy and single-epoch virial mass 
estimators, and galaxy total stellar masses are calculated by fitting 
the spectral energy distribution (subtracting the BLAGN component). 
The mass-growth rates of both the black hole and galaxy are 
calculated using \textit{Chandra}/\textit{XMM-Newton} X-ray and 
\textit{Herschel} far-infrared data, reliable measures of the 
BLAGN accretion and galaxy star formation rates, respectively. 
We use Monte Carlo simulations to account for biases in our 
sample, due to both selection limits and the steep slope of the 
massive end of the galaxy stellar-mass distribution. We find our 
sample is consistent with no evolution in the $M_{\rm{BH}}-M_{\ast}$ 
relation from $z\sim 2$ to $z\sim 0$. BLAGNs and their host galaxies 
which lie off the black hole mass $-$ galaxy total stellar mass relation 
tend to have evolutionary vectors anti-correlated with their mass 
ratios: that is, galaxies with over-massive (under-massive) black holes 
tend to have a low (high) ratio of the specific accretion rate to the 
specific star formation rate. We also use the measured growth rates to 
estimate the preferred AGN duty cycle for our galaxies to evolve 
most consistently onto the local $M_{\rm{BH}}-M_{\rm{Bul}}$ relation. 
Under reasonable assumptions of exponentially declining star formation 
histories, the data suggest a non-evolving (no more than a factor of 
a few) BLAGN duty cycle among star-forming galaxies of $\sim 10\%$ 
($1\sigma$ range of $1-42\%$ at $z<1$ and $2-34\%$ at $z>1$). 
\end{abstract}


\section{Introduction}
\label{sec:intro}
As the most luminous persistent sources in the Universe, active 
galactic nuclei (AGNs), which are powered by the mass accretion 
of super-massive black holes (SMBHs), are widely believed to play 
important roles in the formation and evolution of typical massive 
galaxies \citep[for a recent review, see][]{kor13}. Indeed, over 
the past sixteen years, many studies have revealed that there are 
tight correlations between SMBHs and the physical properties of 
host galaxies \citep[e.g.,][]{mag98,fer00,geb00,tre02,mar03,hr04, 
gul09}. For example, the SMBH mass, $M_{\rm BH}$, correlates well 
with the stellar mass of the bulge, $M_{\rm Bul}$. The intrinsic 
scatter of the $M_{\rm BH}-M_{\rm Bul}$ relation is found to be 
$\sim 0.3\ \rm{dex}$. Such a tight correlation might be explained 
by AGN feedback \citep[e.g.,][]{sil98, kin03, fab12}. In this 
scenario, winds or jets launched from the accretion disk of the 
SMBH heat the interstellar medium (ISM) or clear the ISM out of the 
host galaxy. Star formation is therefore quenched due to the lack of 
cold gas. Such a coupled AGN triggering and feedback-regulated star 
formation could be responsible for the $M_{\rm BH}-M_{\rm Bul}$ 
relation \citep[see, e.g.,][]{dmt05, hop06}. On the other hand, a 
non-causal origin of the $M_{\rm BH}-M_{\rm Bul}$ relation has also 
been proposed by \cite{jah11} \citep[see also][]{peng07}. In this 
scenario, $M_{\rm BH}$ and $M_{\rm Bul}$ are not initially 
correlated. The $M_{\rm BH}-M_{\rm Bul}$ relation is established 
simply by galaxy-galaxy mergers via the central limit theorem. 
Direct observational measurement of the evolution of the 
$M_{\rm BH}-M_{\rm Bul}$ relation is crucial to investigate 
the formation and co-growth of SMBHs and their hosts. 

Observationally, the evolution of the $M_{\rm BH}-M_{\rm Bul}$ 
relation is still a controversial topic \citep[e.g.,][]
{peng06,tre07,jah09,mer10,ben11,sch13}. Moreover, at high 
redshift (e.g., $z\sim 1$), it is very difficult to separate 
the bulge from the total stellar mass. Instead, most works 
explore the $M_{\rm BH}-M_{\ast}$ relation, where $M_{\ast}$ is 
the galaxy total stellar mass. For instance, \cite{jah09} find 
that at $z\sim 1.4$ the $M_{\rm BH}-M_{\ast}$ relation is 
consistent with the local $M_{\rm BH}-M_{\rm Bul}$ relation. Note 
that the host galaxies of the local sample of \cite{hr04} (hereafter 
HR04) are bulge dominated (i.e., $M_{\rm Bul} \sim M_{\ast}$). 
Therefore, the \cite{jah09} result suggests little evolution of the 
$M_{\rm BH}-M_{\ast}$ relation to $z\sim 1.4$. On the other hand, 
the hosts of the \cite{jah09} sample have substantial disks (i.e., 
$M_{\rm{Bul}}<M_{\ast}$). That is, the $M_{\rm BH}-M_{\rm Bul}$ 
relation does evolve as a function of redshift. Making the local 
$M_{\rm BH}-M_{\rm Bul}$ relation then, as pointed out by \cite{jah09}, 
requires a redistribution of stellar mass from disks into bulges in 
AGN host-galaxies. Recently, the work of \cite{sch13} with a larger 
and less-biased sample also supported the results of \cite{jah09}. 
On the other hand, \cite{mer10} compiled a broad emission-line AGN 
(BLAGN\footnote{Throughout this work, we use ``BLAGN'' and ``AGN'' 
interchangeably.}) sample with well-measured multi-band photometry 
from the COSMOS survey. With this sample, they find that even the 
$M_{\rm BH}-M_{\ast}$ relation evolves (at $5\sigma$ significance 
level) over cosmic time \citep[i.e., the SMBH is over-massive at 
high redshifts; see also][]{ben11}. However, as pointed out by \cite{lau07}, 
high-redshift samples are generally selected by AGN activity (which 
depends on $M_{\rm BH}$) and suffer from Eddington bias. The physical 
reasons are: (1)\ there is intrinsic scatter in the 
$M_{\rm BH}-M_{\rm Bul}$ or $M_{\rm BH}-M_{\ast}$ relation; (2)\ the 
galaxy stellar mass function is bottom heavy (i.e., the number 
density of galaxies steeply decreases to higher stellar mass). Hence, 
for these high-redshift samples, an AGN with given $M_{\rm BH}$ is 
more likely to be found in less massive galaxies, resulting in an 
apparent evolution of the $M_{\rm{BH}}-M_{\rm{Bul}}$ or 
$M_{\rm BH}-M_{\ast}$ relation (also see Section~\ref{sec:bias}). 
This bias increases with redshift as at high redshifts we often only 
have access to very luminous AGNs (very massive SMBHs). Additionally, 
$M_{\rm BH}$ estimation itself is also biased for these high-redshift 
samples \citep{shk10}. The reasons are similar: (1)\ $M_{\rm BH}$ is 
estimated from AGN luminosity which also has intrinsic scatter; (2)\ 
the AGN luminosity function is also bottom heavy. Note that the sample 
of \cite{sch13} consists of mostly lower luminosity AGNs and hence 
suffers less bias than \cite{mer10}. Bias corrections must be made to 
compare the high-redshift $M_{\rm BH}-M_{\ast}$ relation with the local 
one \citep[for a more recent discussion on this topic, see, e.g.,][]{sw14}.

In addition to measuring the $M_{\rm BH}-M_{\ast}$ relation at high 
redshift, it is also valuable to study its instantaneous evolution; 
i.e., the SMBH and galaxy stellar mass growth rates. Especially, 
will future SMBH and galaxy mass increases help establish/maintain 
the $M_{\rm BH}-M_{\ast}$ relation? This question has been indirectly 
investigated. For example, many works explored the 
possible connection between star formation rates (SFRs) and AGN 
luminosities \citep[e.g.,][and also see Alexander \& Hickox 2012 for a 
review]{raf11,mul12,ros12,ros13a,ros13b,che13}. However, as pointed out 
by \cite{hic14}, AGN variability can dilute the intrinsic 
correlation between star formation and AGN activity. Therefore, this 
connection is expected to be observed only in an average sense, as 
individual galaxies will tend to show a much weaker (or absent) correlation. 
The average SMBH accretion rate traces the average SFR so well that the 
local $M_{\rm BH}-M_{\ast}$ relation can be established. 

Simultaneously exploring both the $M_{\rm BH}-M_{\ast}$ relation 
at high redshifts and its instantaneous evolution is even more 
valuable but is challenging. \cite{mer10} explore this issue 
by measuring $M_{\rm BH}$, $M_{\ast}$, and their growth rates. 
They find that star formation and AGN activity can actually help 
reduce the scatter of the $M_{\rm BH}-M_{\ast}$ relation. 
However, the SFR in their work is estimated by fitting a stellar 
population synthesis model to the optical-to-NIR spectral energy 
distribution (SED) which has an uncertainty of $\sim 0.7\ \rm{dex}$. 
A more accurate determination of SFR (e.g., with \textit{Herschel} 
observations) could better elucidate the coupled growth of AGNs and 
their host galaxies.

In this work, we uniquely explore the $M_{\rm{BH}}-M_{\ast}$ 
relation and its evolution, accounting for biases and using 
robust and uncontaminated estimates of SMBH growth and 
galaxy star formation rate. In Section~\ref{sec:data}, we 
describe our sample in detail. In Section~\ref{sec:property}, 
we describe our methods for estimating $M_{\rm BH}$, $M_{\ast}$, 
$\dot{M}$, and $\rm SFR$. In Section~\ref{sec:result} we discuss 
selection biases and present our results. In Section~\ref{sec:dis}, 
we discuss implications of our results. Finally, we summarize our 
main results and outline the possible future improvements in 
Section \ref{sec:summ}. Throughout this work we adopt a flat 
$\Lambda$CDM cosmology with $H_0=70\ \rm{km\ s^{-1}\ Mpc^{-1}}$, 
$\Omega_{\rm{M}}=0.3$ and $\Omega_{\Lambda}=0.7$.

\section{Sample Construction}
\label{sec:data}
The $2$-$\rm{deg}^2$ Cosmic Evolution Survey \citep[COSMOS,][]{sco07} 
and $464.5$-$\rm{arcmin}^2$ Chandra Deep Field-South survey 
\citep[CDF-S; e.g.,][]{luo08, luo10, xue11} both have well-calibrated 
multi-band photometry (from the X-ray to radio bands) and high-quality 
spectroscopic coverage. Hence, we selected our sample from 
the COSMOS and CDF-S fields. First, we selected BLAGNs based on 
X-ray detection of AGN activity \citep[for the advantages of using 
X-ray selection, see][]{nba10} and subsequent optical spectroscopic 
identification of broad emission lines (with coverage of either 
$\rm{H}\beta$ or \MgII). Second, we cross-matched our BLAGNs with 
the \textit{Herschel} 
PEP\footnote{\url{http://www.mpe.mpg.de/ir/Research/PEP/index}} 
\citep[][]{lut11} \citep[for the CDF-S field, we also used 
the GOODS--\textit{Herschel} catalogs, see][]{elb11, mag13} and 
HerMES\footnote{\url{http://hedam.lam.fr/HerMES/}} 
\citep[][]{oli12,wan13} catalogs. Third, the new sample was 
cross-matched with existing multi-band photometric catalogs to 
build the final sample. In the following subsections, we discuss 
our sample selection in detail (see also Table 1).

\subsection{COSMOS Field BLAGNs}
The $2$-$\rm{deg}^2$ COSMOS field has been surveyed in the X-ray by 
\textit{XMM-Newton} \citep{cap09, bru10} and 
partially\footnote{This together with the on-going Chandra 
COSMOS Legacy Survey will cover the full COSMOS field \citep{civ13}.} 
(0.9 $\rm{deg}^2$) by \textit{Chandra} \citep{elv09, civ12}. We first 
selected all $z<2.4$ BLAGNs (ensuring the coverage of either 
$\rm{H}\beta$ or \MgII) from the public COSMOS spectroscopy 
\citep{lil07,tru09a}. Our BLAGNs have public 
Magellan/IMACS \citep{tru09a}, SDSS \citep{yor00}, or VLT/VIMOS 
\citep[zCOSMOS;][]{lil07} spectroscopic 
data\footnote{\url{http://irsa.ipac.caltech.edu/Missions/cosmos.html}} 
which allow us to estimate $M_{\rm BH}$ via the single-epoch approach 
(see also Section~\ref{sec:mbh}). The Magellan/IMACS spectroscopy is 
limited by $i_{\rm AB}<22.5$ \citep{tru09a}. For consistency and ease of 
modeling the observational biases, we impose the same $i_{\rm AB}<22.5$ 
limit for the combined COSMOS and CDF-S sample. We then cross matched 
them with X-ray catalogs to get the corresponding (rest-frame) 
$2-10\ \rm{keV}$ luminosity. For X-ray observations, we prefer \textit{Chandra} 
over \textit{XMM-Newton} and hard band over soft band (i.e., the \textit{Chandra} 
hard band has the highest priority, and the \textit{XMM-Newton} soft band has 
the lowest priority). As pointed out by \cite{tru09a, tru11} and will be shown in 
Figure~\ref{fig:bolz}, our AGN selection is limited by the $i_{\rm AB}<22.5$ 
criteria and not the X-ray sensitivity (whether \textit{XMM-Newton} or 
\textit{Chandra}).

To get a robust estimate of SFR, we cross-matched our BLAGN sample 
with the \textit{Herschel} PEP ($100\ \rm{\mu m}$ and $160\ \rm{\mu m}$) 
and HerMES ($250\ \rm{\mu m}$) catalogs. Note that for the HerMES 
catalog the SPIRE/\textit{Herschel} observations have a large point-spread 
function. We followed \cite{che13} and chose a maximum matching radius 
of $5^{''}$. The \textit{Herschel} PEP catalogs have improved astrometry 
because sources were extracted using \textit{Spitzer}/MIPS $24\ \rm{\mu m}$ 
priors, and so we used a maximum matching radius of $2^{''}$.\footnote{For 
the \textit{Herschel} catalogs, our maximum matching radii lead to a 
false match rate of $1.3\%$ for PEP and $9\%$ for HerMES. (All but five 
objects have both PEP and HerMES data.) The false-match rate is calculated 
by following the procedure described in Section~3 of \cite{luo08}.} 
As for multi-band photometry, we used the COSMOS Photometric Redshift 
Catalog Fall 2008 \citep{llb09} to get the optical-to-NIR broad-band 
photometry information ($B$, $V$, $g$, $r$, $z$ of Subaru; $u$, $i$, 
$Ks$ of CFHT; and $J$ of UKIRT) for the selected sources \citep[see 
also][]{cap07, mcc10}. We also included four \textit{Spitzer}/IRAC 
channels \citep{san07} and \textit{GALEX}/Near UV data when available. 
Zero-point corrections suggested in \cite{llb09} were applied. The 
Galactic extinction was also corrected using the Galactic extinction 
map of \cite{sfd98}. Note that for the optical-to-NIR bands, the 
Galactic extinction corrections are already included in the catalog. 

\subsection{CDF-S Field BLAGNs}
The $464.5$-$\rm{arcmin}^2$ CDF-S field is covered by the deepest 
X-ray survey, with $4$ Ms of \textit{Chandra} coverage \citep{xue11}. 
We selected BLAGNs by cross-matching the 4 Ms CDF-S catalog with the 
optical spectroscopy of \cite{szo04}.\footnote{\url{http://www.mpe.mpg.de/CDFS/data/}} 
Similar to the COSMOS sources, we only selected BLAGNs (with 
$i_{\rm AB}<22.5$) whose host galaxies are also detected by the 
\textit{Herschel} PEP or HerMES (the cross-matching criteria is the 
same as that of COSMOS BLAGNs). As for multi-band photometry, 
we used the MUSYC\footnote{\url{http://www.astro.yale.edu/MUSYC/}} 
32-band (including $U$, $U38$, $B$, $V$, $R$, $I$, $z$, $J$, $H$, $K$ 
bands; 18 Subaru medium bands; and four \textit{Spizter}/IRAC channels) 
catalog \cite[][and references therein]{car10}. The Galactic extinction 
and zero-point corrections suggested in \cite{car10} were adopted. 

\subsection{Total BLAGNs}
Based on the above criteria, we started with a total number of 
251 X-ray BLAGNs at $z<2.4$ (ensuring coverage of either 
$\rm{H}\beta$ or \MgII), with 221 from the COSMOS field and 
30 from the CDF-S field. When cross-matched with the 
\textit{Herschel} catalogs, 96 sources were left (85 
from the COSMOS field; 11 from the CDF-S field). 
69 of these 96 sources have high-quality spectroscopic 
observations and robust FIR detections which enable us to measure 
$M_{\rm BH}$, $M_{\ast}$, $\dot{M}$, and SFR, reliably. For the other 
27 sources, either the spectroscopic data (nine objects) 
or the FIR observations (18 objects) are not sufficient for a reliable 
measurement of $M_{\rm BH}$ or SFR (see Sections~\ref{sec:mbh} \& 
\ref{sec:sfr}). Note that due to the \textit{Herschel} sensitivity, $\sim 2/3$ 
of BLAGNs are undetected. We will consider the bias introduced by the 
\textit{Herschel} sensitivity in Section~\ref{sec:bias}. 

For the 69 sources in our sample, 62 are from the COSMOS field and 
seven are from the CDF-S field. Note that their spectra are flux 
calibrated. We compared the physical properties 
(Section~\ref{sec:property}) of these seven CDF-S sources with the 
COSMOS sources (using the Mann-Whitney U test) and found no statistical 
difference (i.e., the null probability $p>0.05$).

\section{Measuring Properties of AGNs and Their Host Galaxies}
\label{sec:property}
\subsection{Black-Hole Mass Estimation}
\label{sec:mbh}
We adopted single-epoch virial SMBH mass estimators to estimate SMBH 
masses. More specifically, H$\beta$ or \MgII\ line estimators were 
used depending on the redshift. These SMBH mass 
estimators use the correlation between the radius of the broad line 
region (BLR), $R_{\rm BLR}$, and the continuum luminosity, 
$\lambda L_{\lambda}$, $R_{\rm{BLR}}\propto (\lambda L_{\lambda})^{0.5}$, 
revealed by reverberation mapping observations of local AGNs 
\citep{ben06,kas07}. If the BLR is virialized, then the SMBH mass can be 
estimated by $M_{\rm{BH}}=fR_{\rm{BLR}}v^2_{\rm{FWHM}}$, where 
$v_{\rm{FWHM}}$ is the full width at half maximum of the broad emission 
line and $f$ is related to the BLR geometry \citep[calibrated from 
dynamical estimates of $M_{\rm BH}$, see, e.g.,][]{onk07}. Therefore, 
we can estimate the SMBH mass from $\lambda L_{\lambda}$, and $v_{\rm{FWHM}}$, 
\begin{equation}
\label{eq:mse}
\log (\frac{M_{\rm{BH}}}{M_{\odot}}) = A + B\log (\lambda L_{\lambda}) 
+ 2\log (v_{\rm{FWHM}}) \\, 
\end{equation}
where $\lambda L_{\lambda}$ and $v_{\rm{FWHM}}$ are in units 
of $10^{44}\ \rm{erg\ s^{-1}}$ and $1000\ \rm{km\ s^{-1}}$, respectively. 
This relation has an intrinsic scatter of $\sim 0.4$ dex \citep{ves06}, 
which is due to the unknown BLR geometry and other factors (e.g., 
ionization state). Note that for H$\beta$, $\lambda = 5100\ \rm{\AA}$, 
$A=6.91$, $B=0.5$ \citep{ves09}; for \MgII, $\lambda = 3000\ \rm{\AA}$, 
$A=6.86$, $B=0.47$ \citep{ves06}. Both $\lambda L_{\lambda}$ and 
$v_{\rm{FWHM}}$ can be measured by fitting the single epoch spectra. 

We performed an iterative 
chi-squared minimization\footnote{Using the \emph{lmfit} python package, 
available from \url{http://cars9.uchicago.edu/software/python/lmfit/index.html}} 
to fit the broad emission lines (H$\beta$ or \MgII) for each source 
and measured $v_{\rm{FWHM}}$ and $\lambda L_{\lambda}$. The 
fitting procedures were similar to those of \cite{tru09b}. That is, a 
pseudo-continuum \citep[power-law continuum plus broadened Fe template 
of][]{ves01} and one or two broad Gaussian line profiles were 
(simultaneously) used to get a best fit of each spectrum. For H$\beta$, 
we also added [O\,{\sc iii}]\,$\lambda\lambda 4959,5007$ lines and a 
narrow H$\beta$ component to the fit and then subtracted them when 
fitting the broad emission lines since these narrow components come 
from the narrow-line region. Our criterion for the 
use of one or two broad Gaussian line profiles was as follows: first, 
we used two Gaussian components to fit the broad emission line and 
then we compared $v_{\rm{FWHM}}$ of both Gaussian components. If at 
least one component had $v_{\rm{FWHM}}<10^3\ \rm{km\ s^{-1}}$, we 
refitted the spectrum with only one Gaussian component. Note 
that if we use two Gaussian line profiles for all sources, our 
$M_{\rm BH}$ estimates change by no more than $0.2$ dex, and this 
is small compared to the intrinsic uncertainty of $M_{\rm{BH}}$. 

\begin{figure}
\plotone{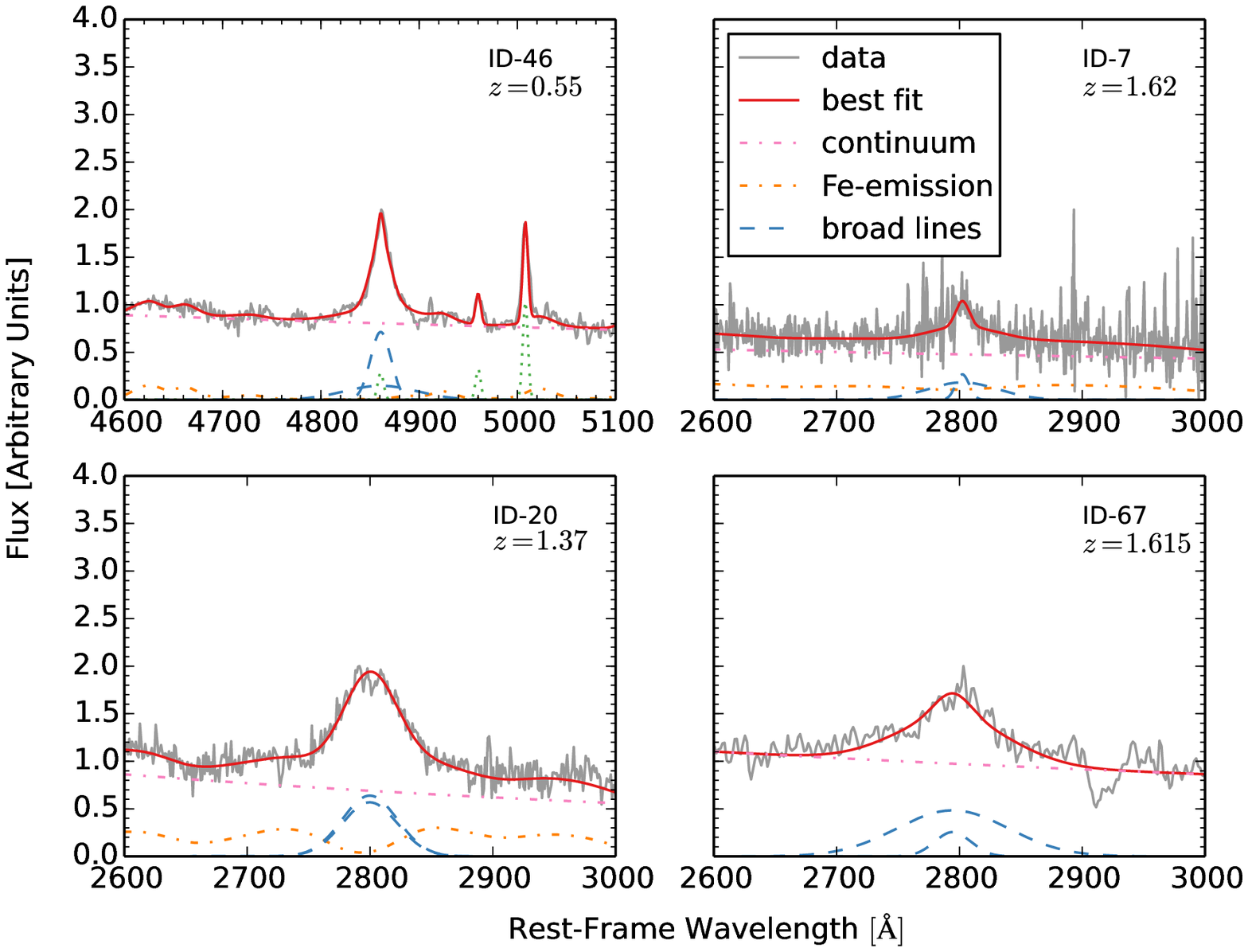}
\caption{
Examples of multi-component fits to the broad H$\beta$ (the upper-left 
panel) or \MgII\ emission lines. Dotted lines represent minor features 
(e.g., the [O\,{\sc iii}]\,$\lambda\lambda 4959,5007$ lines and the 
narrow H$\beta$ component which are removed in the broad-line fit). 
}
\label{fig:fbh}
\end{figure}

In Figure~\ref{fig:fbh}, we present examples 
of our fits to the broad emission lines. The gray and red solid lines 
represent the observed spectra and our best fit, respectively. Red 
and orange dot-dashed lines represent the power-law continuum 
component and Fe template, respectively. The dotted lines correspond 
to minor features (i.e., [O\,{\sc iii}]\,$\lambda\lambda 4959,5007$ 
lines, the narrow H$\beta$ line, which are excluded from the broad-line 
fit). Blue dashed lines represent best fits of broad emission lines. We 
calculated $v_{\rm{FWHM}}$ as follows: if two broad Gaussian profiles 
were used, we calculated $v_{\rm{FWHM}}$ from the emission profile which 
is the summation of the two broad Gaussian components; otherwise, we 
calculated $v_{\rm{FWHM}}$ from only the one broad Gaussian profile. 
We have visually inspected all fitting results and found nine 
sources whose spectroscopic line profiles do not allow reliable 
measurement of $M_{\rm{BH}}$.\footnote{Four 
of these nine sources only show narrow H$\beta$ lines in their spectra; i.e., 
they are misclassified as BLAGNs, and are instead Type 2 or intermediate 
type (type 1.x) AGNs. Five of these nine sources show strong absorption 
features superposed on the broad emission lines. The removal of these 
sources should not introduce a selection bias to our sample, since they 
are essentially random occurrences which are unlikely to be strongly 
correlated with Eddington ratio or host properties.} We excluded these 
nine sources from our sample. Meanwhile, $\lambda L_{\lambda}$ was 
calculated from the power-law continuum component. Note that we also 
applied a small correction to account for the contamination of the stellar 
light of the host galaxy suggested in Section~\ref{sec:mgl}. 

A summary of SMBH masses can be found in Table~2 (before compiling the 
Table, we removed another 18 sources because of inadequate 
FIR data, see Section~\ref{sec:sfr}). There are 24 (46) sources 
in Table 2 whose $M_{\rm{BH}}$ were measured using H$\beta$ (\MgII). 
Previous studies \citep[e.g.,][]{shl12} have revealed that the two 
$M_{\rm{BH}}$ estimators are consistent. Four sources in Table 2 have 
spectroscopic data enabling us to measure $M_{\rm{BH}}$ using both 
H$\beta$ and \MgII. For three BLAGNs, the two $M_{\rm{BH}}$ estimators 
are consistent with each other within $0.2$ dex. For the remaining BLAGN 
(ID-48), the two $M_{\rm{BH}}$ estimators show a larger deviation ($0.6$ 
dex) but are still consistent within the $2\sigma$ uncertainty. In any 
case, we preferred the H$\beta$ estimator over \MgII\ estimator.

\subsection{Galaxy Total Stellar Mass Estimation}
\label{sec:mgl}
We utilized an SED fitting technique to estimate $M_{\ast}$. Since 
we were dealing with BLAGNs, the SEDs (from UV to Near IR) of the host 
galaxies are contaminated by AGN emission. We therefore followed the 
approach of \cite{bon12} \citep[see also][]{mer10} which fits the 
observed SEDs as a composite of both AGN and galaxy stellar emission. 
That is, 
\begin{equation}
f_{\rm{obs}, \lambda} = c_{\rm{AGN}}f_{\rm{AGN}, \lambda} + 
c_{\rm{Gal}}f_{\rm{Gal}, \lambda} \\,
\end{equation} 
where $f_{\rm{AGN}, \lambda}$ and $f_{\rm{Gal}, \lambda}$ are the 
fluxes from the AGNs and host galaxies, respectively. For SEDs, 
fourteen bands (observed-frame wavelength ranging from $2500\ \rm{\AA}$ 
to $79595\ \rm{\AA}$) are used for the COSMOS sources and 32 bands 
(observed-frame wavelength ranging from $3656\ \rm{\AA}$ to $79595\ \rm{\AA}$) 
are used for the CDF-S sources. For AGN emission, we adopted the mean 
SED template of \cite{R06}. For galaxy 
templates, we selected from an SED library constructed using the 
\cite{bc03} stellar population synthesis model. We assumed an initial 
mass function (IMF) of \cite{imf} and adopted ten exponentially 
declining star formation histories (SFHs), i.e., SFR 
$\propto\ e^{-t_{\rm age}/\tau}$, with star formation timescale $\tau$ 
ranges from $0.1$ to $30\ \rm{Gyr}$, plus a SFH with constant SFR. 
The galaxy age ranges from $50\ \rm{Myr}$ to $9\ \rm{Gyr}$ with an 
additional constraint that the galaxy age should not be larger than 
the age of the Universe at the redshift of the source. The SED library 
was generated iterating over $\tau$ and the galaxy age. When fitting 
the observed SED, we also took intrinsic extinction into consideration. 
For AGNs, we used a Small Magellanic Cloud-like dust-reddening curve 
\citep{pre84} with $E(B-V)\leq 1.0$. For galaxies, we adopted the 
Calzetti extinction curve \citep{cal00} with $E(B-V)\leq0.5$ if 
$t_{\rm{age}}/\tau<4$, otherwise $E(B-V)\leq 0.15$ 
\citep{fon06,poz07,bon12}. 

We additionally required that the AGN continuum emission should be 
larger than $10\%$ of the galaxy continuum emission at $2800\ \rm{\AA}$ 
(if the \MgII\ virial mass estimator was used) or $4861\ \rm{\AA}$ (if 
the H$\beta$ virial mass estimator was used). This was simply motivated 
by the fact that our sources all have observed broad emission lines 
(H$\beta$ or \MgII) indicative of a significant AGN contribution. We 
chose $10\%$ based on the following considerations: (1)\ we performed 
a simple simulation by creating a mock spectrum with a young galaxy 
SED, an AGN (power-law) component, a broad Gaussian emission 
line with a typical equivalent width of $30\ \rm{\AA}$ ($60\ \rm{\AA}$) 
for \MgII\ (H$\beta$) \citep[e.g., Section 1.3.4 of][]{agnbook}, and white 
noise (assuming a signal-to-noise ratio of $\sim 10$); (2)\ we refitted 
the mock spectrum with a power-law and a Gaussian emission line and 
found that we cannot reliably fit the emission line if the AGN emission is 
less than $\sim 10\%$ of galaxy emission. Furthermore, we have performed 
a Monte Carlo simulation to mimic our selection procedures (see 
Section~\ref{sec:bias} for more details). For our simulated sample, 
we also found that there are a negligible number of sources with AGN 
emission less than $10\%$ of the galaxy emission at $2800\ \rm{\AA}$ 
(or $4861\ \rm{\AA}$).

\begin{figure}
\plotone{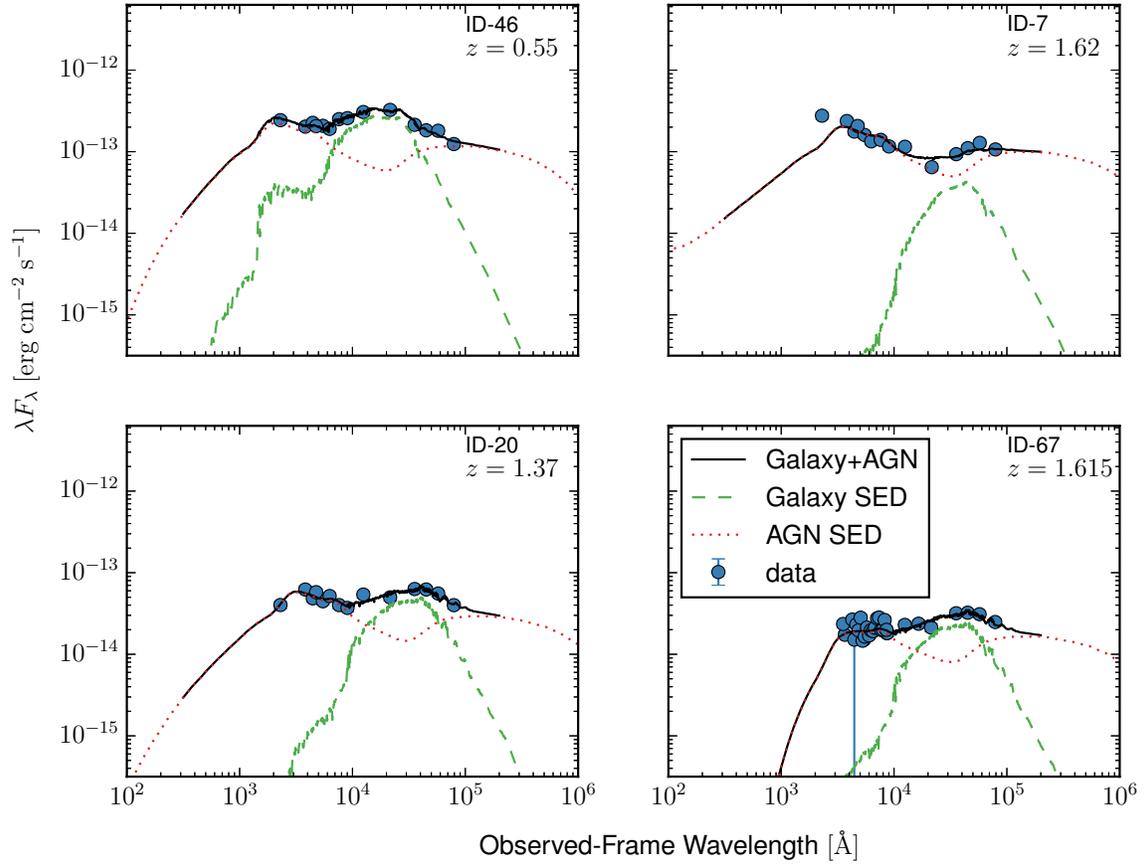}
\caption{
Examples of two-component (AGN+galaxy) fits to SEDs (from the NUV to 
NIR). Red dotted (green dashed) lines represent the AGN (galaxy) 
component. Black solid lines show the summation of the galaxy and AGN 
SEDs. Note that each data point has a photometric error which is 
typically too small to be visible on this scale. 
}
\label{fig:fgm}
\end{figure}

\begin{figure}
\plotone{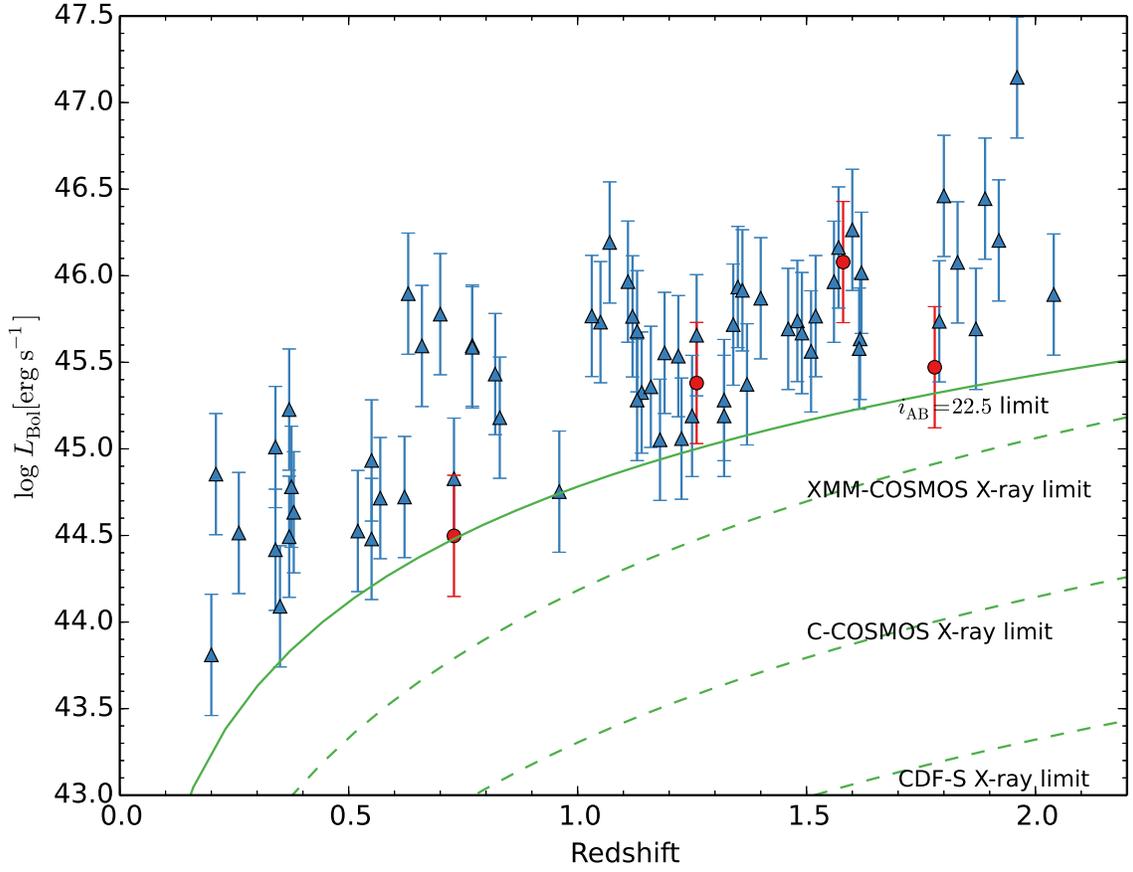}
\caption{
The bolometric luminosity of AGNs (calculated by combining the X-ray 
observations and SED fitting) vs. redshift. Blue triangles represent 
AGNs that are detected in the observed-frame hard X-ray band 
($2-10\ \rm{keV}$); red points are sources that are detected in the 
observed-frame soft X-ray band ($0.5-2\ \rm{keV}$) but not in the 
observed-frame hard X-ray band. The green solid line represent the 
limit introduced by $i_{\rm{AB}}=22.5$ which corresponds to the 
$\sim 90\%$ completeness of the optical spectroscopy for COSMOS 
sources \citep{tru09a}. The bolometric correction of the $i$ band is 
assumed to be $12$ \citep{R06} and the $K$ correction is made 
by assuming the optical-to-UV SED is a power-law with 
$\alpha_{\nu}=-0.44$ \citep{van01}. 
The three green dashed lines represent the limits introduced by the X-ray 
sensitivity of the soft band of \textit{XMM-Newton}/COSMOS and 
\textit{Chandra}/COSMOS and the hard band of CDF-S, respectively. 
The bolometric correction of \cite{hop07} is assumed to 
calculate $L_{\rm Bol}$ from the X-ray sensitivity. 
}
\label{fig:bolz}
\end{figure}

We fitted $f_{\rm{obs}, \lambda}$ to the observed SEDs by performing 
iterative (reduced) chi-squared minimization. Figure~\ref{fig:fgm} shows 
examples of our two-component (AGN plus galaxy) fitting to the observed 
SEDs. The data are well described by a combination of AGN and galaxy 
emission. We also inspected the corresponding HST/ACS images and found that 
generally our SED fits are consistent with the morphology information. 
That is, when the SED fit suggests the AGN emission dominates in the $i$ 
band, the HST/ACS image also shows a point-like morphology (the inverse 
is also true). From each best fit, we determined both the normalization, 
$t_{\rm age}$, $\tau$, and the SFH of our galaxy SED. We then used them 
and the \cite{bc03} model to calculate the galaxy total stellar mass. A 
summary of the galaxy total stellar masses can be found in Table 2. Note 
that the two-component SED fit can also tell us the AGN contribution to 
either the $3000\ \rm{\AA}$ or $5100\ \rm{\AA}$ luminosity, and we 
can use this information to calculate the true AGN contribution to 
$\lambda L_{\lambda}$ measured in Section~\ref{sec:mbh}. This small 
correction (accounting for the contamination of the stellar light of 
the host galaxy) has been included when calculating $\lambda L_{\lambda}$ 
from the continuum component in Section~\ref{sec:mbh}. 

\subsection{Accretion-Rate Estimation}
\label{sec:acc}
To measure the accretion rate, $\dot{M}$, we estimated the bolometric 
luminosity, $L_{\rm{Bol}}$ \citep[e.g.,][]{sol82}. The accretion rate is, 
\begin{equation}
\label{eq:mdot}
\dot{M} = \frac{(1-\eta)L_{\rm{Bol}}}{\eta c^2} \\,
\end{equation}
where $\eta=0.1$ is the assumed radiative efficiency of the accretion 
disk, and $c$ is the speed of light. We used two methods to calculate 
$L_{\rm{Bol}}$: (1)\ the rest-frame $2-10\ \rm{keV}$ luminosity 
alone; (2)\ the SED fit plus the rest-frame $2-10\ \rm{keV}$ 
luminosity. 

We first used the rest-frame $2-10\ \rm{keV}$ luminosity as an 
estimator of $L_{\rm{Bol}}$. The rest-frame $2-10\ \rm{keV}$ 
luminosity is calculated from the observed-frame $2-10\ \rm{keV}$ 
flux (assuming a typical power-law photon index of $\Gamma=1.7$). 
Note that there are four sources in our sample that are undetected 
in the observed-frame $2-10\ \rm{keV}$ band. For them, we instead 
calculated the rest-frame $0.5-2\ \rm{keV}$ luminosity from the 
observed-frame $0.5-2\ \rm{keV}$ flux (again with $\Gamma=1.7$) and 
used (different) bolometric corrections to estimate $L_{\rm{Bol}}$. 
We adopted the \cite{hop07} luminosity-dependent bolometric 
corrections:
\begin{equation}
\kappa_{\rm{Band}} = \frac{L_{\rm{Bol}}}{L_{\rm{Band}}} = 
d_1(\frac{L_{\rm{Bol}}}{10^{10}L_{\odot}})^{k_1} + 
d_2(\frac{L_{\rm{Bol}}}{10^{10}L_{\odot}})^{k_2} \\,
\end{equation}
where $\kappa_{\rm{Band}}$ is the bolometric correction factor. 
The constants $(d_1, k_1, d_2, k_2)$ are given by 
$(17.87, 0.28, 10.03, -0.02)$ for $L_{0.5-2\ \rm{keV}}$, and 
$(10.83, 0.28, 6.08, -0.02)$ for $L_{2-10\ \rm{keV}}$. The 
uncertainty of $\kappa_{\rm{Band}}$ is \citep{hop07}, 
\begin{equation}
\sigma_{\log(\kappa)}=\sigma_1(L_{\rm{Bol}}/10^9L_{\odot})^{\beta}+
\sigma_2 \\,
\end{equation}
with ($\sigma_1$, $\beta$, $\sigma_2$) $= (0.046, 0.10, 0.08)$ in 
the $0.5-2\ \rm{keV}$ band and $= (0.06, 0.10, 0.08)$ in the $2-10\ 
\rm{keV}$ band. For sources in our sample, the median of the 
uncertainties is $\sim 0.2$ dex. 

We also combined the SED fitting results presented in Section~\ref{sec:mgl} 
with X-ray observations to calculate the bolometric luminosity. 
This procedure was similar (but not identical) to that of \cite{tru11}. 
First, we used the rest-frame $2-10\ \rm{keV}$ luminosity to get the 
$0.5-250\ \rm{keV}$ luminosity (assuming $\Gamma=1.7$ and a cut-off 
at $250\ \rm{keV}$); second, we integrated the big blue bump component 
(from $30\ \rm{\AA}$ to $10^4\ \rm{\AA}$) in the best-fitted AGN SED which 
should be responsible for the emission from the accretion disk (note that we 
neglected the IR ``torus'' emission as it is probably reprocessed); 
finally, we obtained the bolometric luminosity by summing the 
$0.5-250\ \rm{keV}$ luminosity and the total big blue bump luminosity. 
This approach gives a direct estimate of $L_{\rm{Bol}}$. 

We compared the two bolometric luminosity estimates and found that 
the ratios of these two estimates follow a log-normal distribution. 
This log-normal distribution has a mean value of $0.03$ and a standard 
deviation of $0.4$ dex. This deviation should be caused by the 
uncertainties of the two bolometric luminosity estimators. Therefore, 
we conclude that the uncertainty of the bolometric luminosity estimated 
using the SED fit plus the X-ray luminosity is 
$(0.4^2-\sigma_{\log(\kappa)}^2)^{0.5}=0.35$ dex. 
In the following calculations, we use the bolometric luminosity 
calculated from SED fits plus X-ray observations to measure $\dot{M}$. 
A summary of the bolometric luminosities can be found in Table 2.

Figure~\ref{fig:bolz} plots $L_{\rm Bol}$ as a function of redshift. 
In addition, we also include the bolometric luminosity limits 
introduced by the $i_{\rm AB}<22.5$ spectroscopy limit and the X-ray 
sensitivity limits of COSMOS (soft bands of \textit{Chandra} and 
\textit{XMM-Newton}) and CDF-S (hard band of \textit{Chandra}). The 
X-ray limits are significantly deeper than the optical spectroscopy limit, 
and the sample is essentially limited only by $i_{\rm AB}<22.5$ 
\citep[see also][]{tru09a}. The four sources which are only detected in 
the observed-frame $0.5-2\ \rm{keV}$ band are denoted as red points. 
The redshifts of our BLAGNs span $0.2\leq z<2.1$ with a median redshift 
of $1.18$. We also define a ``high-redshift'' sub-sample of the 45 sources 
at $z>1$, and a ``low-redshift'' sub-sample of the 25 sources at $z\leq 1$. 

\begin{figure}
\plotone{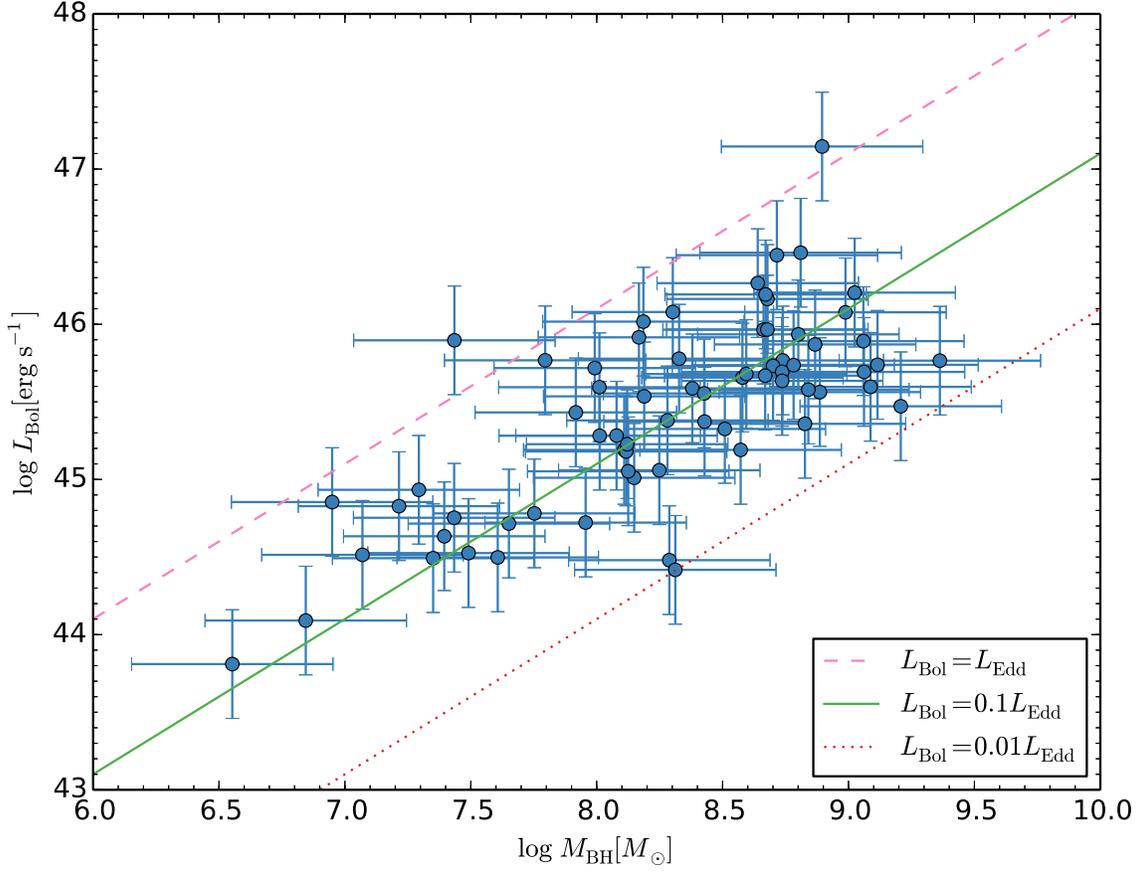}
\caption{
The bolometric luminosity of AGNs vs. $M_{\rm BH}$. The pink dashed, 
green solid and red dotted lines correspond to the Eddington ratios 
of $1.0$, $0.1$ and $0.01$, respectively. For each object, the 
intrinsic scatter in the $L_{\rm Bol}$ and $M_{\rm BH}$ estimators 
($0.35$ and $0.4$ dex, respectively) dominates the error. The Eddington 
ratios of our source spans from $0.01$ to $1.0$, typical range of 
BLAGNs \citep[e.g.,][]{kol06, tru09b, tru11, lus12}.
}
\label{fig:bolmbh}
\end{figure}

Figure~\ref{fig:bolmbh} plots the bolometric luminosity as a function 
of $M_{\rm BH}$. We also include the lines of $L_{\rm Bol}=0.01\ L_{\rm Edd}$, 
$L_{\rm Bol}=0.1\ L_{\rm Edd}$ and $L_{\rm Bol}=L_{\rm Edd}$, where 
$L_{\rm{Edd}}=1.26\times 10^{38}M_{\rm{BH}}/M_{\odot}\ \rm{erg\ s^{-1}}$. 
As clearly seen from this figure, the Eddington ratios 
of our BLAGNs are typically spread from $0.01-1$ \citep[consistent 
with, e.g.,][]{kol06, tru09b, lus12}. For COSMOS sources, we also made 
a detailed comparison between Eddington ratios of our sources and those 
of \cite{tru11} by checking the overlapping sources one by one. We found 
that the median absolute deviation (M.A.D.) of the differences is $0.2$ 
dex, much smaller than the uncertainty we assigned to $M_{\rm BH}$ 
(see Section~\ref{sec:err}). For the CDF-S field, \cite{bab07} found 
a broad distribution of Eddington ratios, using host galaxy mass (or 
velocity dispersion) to estimate SMBH mass. Of their sources with broad 
emission lines, 6/11 have Eddington ratios between $0.01$ and 1, and 
two others have Eddington ratios consistent with $0.01$. Only 3/11 have 
Eddington ratios significantly smaller than $0.01$. Despite the very 
different methods of $M_{\rm BH}$ estimation, our results are broadly 
consistent with their results. 

\subsection{SFR Estimation}
\label{sec:sfr}
Far-IR emission has long been argued to be an excellent SFR tracer 
as it is largely free of dust extinction. We used the widely adopted 
\cite{ken98} relation \citep[with a modification to account for the 
IMF of][]{imf} to derive the SFR:
\begin{equation}
\label{eq:msg}
\mathrm{SFR} (M_{\odot}\ \mathrm{yr^{-1}}) = 1.09\times 10^{-10} 
L_{\rm{IR}}/L_{\odot} \\,
\end{equation}
where $L_{\rm{IR}}$ is the total $8-1000\ \rm{\mu m}$ IR luminosity. 
BLAGNs make significant contributions in the near- to mid-IR bands. 
The FIR band is, however, known to be dominated by galaxy emission 
\citep[e.g., ][hereafter K12]{kir12}. In this work, we used FIR data 
from the \textit{Herschel} PACS/SPIRE bands to normalize the $z\sim 1$ 
star-forming galaxy SED template of K12 and integrated the normalized 
SED to estimate $L_{\rm{IR}}$. There are other popular galaxy IR SED 
templates \citep[e.g.,][]{ce01,dh02}. However, the K12 template is 
directly derived from the high-redshift sources and therefore may be 
more relevant to our high-redshift sources. As stated before, we have 
additionally removed 18 sources which have $z>1$ but are detected only 
at the $100\ \rm{\mu m}$ band. The reason is that, under such 
circumstances, the $100\ \rm{\mu m}$ band might be significantly 
contaminated by AGN emission \citep[e.g., K12,][]{nor12}. Such 
measurements of $L_{\rm{IR}}$ are therefore not reliable. For our 
69 sources, 26 are detected at $100\ \rm{\mu m}$, $160\ \rm{\mu m}$, 
and $250\ \rm{\mu m}$. 24 are detected in two bands. The remaining 19 
sources are detected only in one band. A summary of $L_{\rm{IR}}$ can 
also be found in Table~2. 

\begin{figure}
\plotone{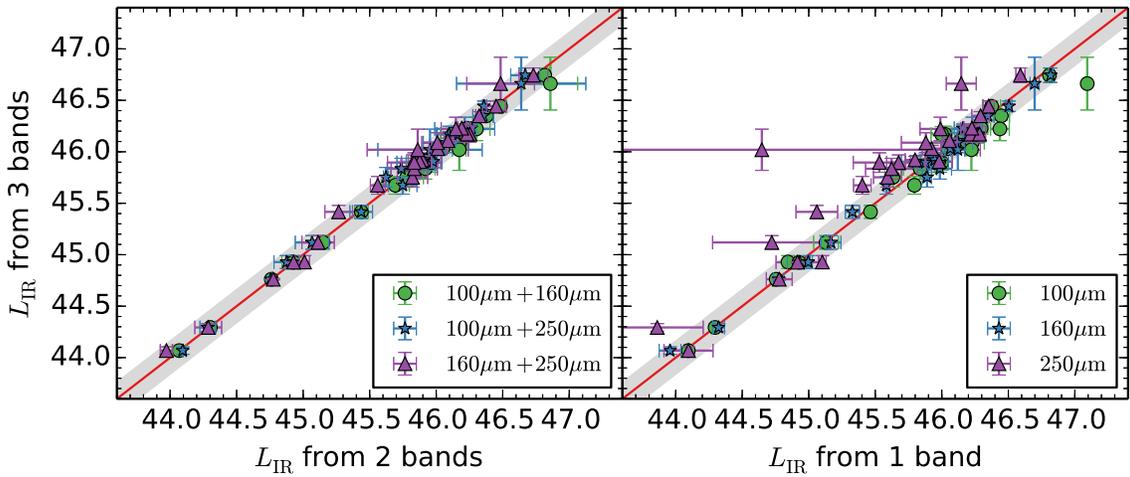}
\caption{
Illustration of the reliability of estimating FIR luminosity using 
different \textit{Herschel} bands. Left: Comparison of FIR luminosity 
estimated from two bands with that from all three bands. Right: 
Comparison of FIR luminosity estimated from one band with that from 
all three bands. The solid line and shaded area represent the ratio 
of unity and $0.17$ dex uncertainties. The two-band FIR estimator 
agrees well with the three-band estimator. Estimating SFR from a 
single band is also consistent with the three-band estimator. Note 
that there is a source in the right panel shows an extremely large 
uncertainty which is caused by the fact that the uncertainty of the 
flux in $250\ \rm{\mu m}$ for this source is large (if we take also 
the $100\ \rm{\mu m}$ and $160\ \rm{\mu m}$ bands into consideration, 
the uncertainty of this source is small). 
}
\label{fig:fira}
\end{figure}

To check the consistency of our $L_{\rm{IR}}$ estimation, we took our 
26 sources that are detected in three bands and compared $L_{\rm{IR}}$ 
values that are estimated from all three bands with those values from 
two bands or one band. Figure~\ref{fig:fira} shows the result. As seen 
from the figure, $L_{\rm{IR}}$ values estimated from three bands are 
consistent with those from only two bands or one band. Therefore, we 
can conclude that our method for estimating $L_{\rm{IR}}$ is 
self-consistent. Note that, even for $z>1$ sources, $L_{\rm{IR}}$ 
estimated from $100\ \rm{\mu m}$ agrees well with that from three 
bands, which indicates that for these three-band detected 
sources AGN emission does not strongly contaminate even the blue 
\textit{Herschel} FIR band. For the removed 18 sources, since 
we only have information on the bluest band (i.e., $100\ \rm{\mu m}$) 
which corresponds to rest-frame $30-50\ \rm{\mu m}$ (for 
$z=1\sim 2$), we cannot rule out the possibility that AGN contamination 
may still be important.\footnote{Note that our results below do not 
change materially if we include the 18 objects by assuming their SFR 
estimation is robust.}

To quantify the star formation activity of our sample, we compared the 
$\mathrm{SFR}-M_{\ast}$ relation of our galaxies with the star-forming 
``main sequence" \citep[e.g.,][]{elb07, noe07, whi12}. The 
$\mathrm{SFR}-M_{\ast}$ relation (i.e., the star-forming 
``main sequence'') we adopted is from \cite{whi12},
\begin{equation}
\label{eq:ms}
\log\ \mathrm{SFR} = a(z)(\log\ M_{\ast}-10.5)+b(z) \\,
\end{equation}
with $a(z)=0.7-0.13z$, $b(z)=0.38+1.14z-0.19z^2$. This relation has 
a scatter of $0.34$ dex, roughly independent of $z$ and $M_{\ast}$. 
Figure~\ref{fig:sfrc} plots the ratio of the SFR we measured to the 
one predicted by Equation~\ref{eq:ms}. For the uncertainty of the 
ratio, we combined the uncertainties of $M_{\ast}$, SFR, and 
the scatter of Equation~\ref{eq:ms}. Our sources show higher star 
formation activity than that of the main sequence (although they 
are not significantly offset by more than their uncertainties; also 
the mean ratio is $\sim 2$, unlike starburst galaxies whose ratios 
have $\gtrsim 3\sigma$ offset from Equation~\ref{eq:ms}). This 
is driven mostly by the \textit{Herschel} flux limit, which lies 
around or above the main sequence \citep[see, e.g., Figure~4 of][]{nor12} 
and causes an Eddington bias (see Section~\ref{sec:bias}). The factor 
of two excess in the mean ratio is fully consistent with the expected 
Eddington bias if the host galaxies of our BLAGNs were drawn 
from the star-forming main sequence. 

\subsection{Error Budget}
\label{sec:err}
In our sample, $M_{\rm{BH}}$, $M_{\ast}$, $\dot{M}$, and SFR all 
have significant uncertainties. Let us first consider the 
uncertainty of $M_{\rm{BH}}$ estimation. For a fraction of our 
sources, $M_{\rm BH}$ has been carefully estimated by some 
previous studies \citep[e.g.,][]{mer10,mat13}. Therefore, we 
compared our $M_{\rm BH}$ estimations with these previous works. 
We found that there is no global systematic offset and the standard 
deviation is no more than $0.3$ dex \citep[actually, when comparing 
with][we found that for all sources but one, the 
deviations are smaller than $0.2$ dex; for the remaining one, the 
deviation is $\sim 0.3$ dex]{mat13}. Therefore, the error is 
dominated by the intrinsic error of the SMBH mass single-epoch 
virial relation. We then adopted $0.4$ dex, the intrinsic error, 
as the error of $M_{\rm{BH}}$ \citep{ves06}. For $\dot{M}$, we only 
considered the uncertainty from $L_{\rm Bol}$, which is $\sim 0.35$ 
dex (see Section~\ref{sec:mgl}).\footnote{For simplicity, in this 
work, we neglect the scatter of $\eta$. Note that our anti-correlations 
in Section~\ref{sec:dis2} do not significantly change if we instead 
assume the scatter of $\eta$ is $\sim 0.5$ dex \citep[e.g.,][]{dav11}}

For $M_{\ast}$, we compared our results with those of \cite{bon12} 
(COSMOS sources) and \cite{xue10} (CDF-S sources). We found the 
median and M.A.D.\ of the differences 
is $0.02$ dex and $0.2$ dex, respectively. Note that there are a 
handful of sources in our sample showing relatively large deviations 
of $M_{\ast}$ with respect to those of \cite{bon12} (although there 
is no global systematic offset). This is likely caused by several 
factors, e.g., different photometric data \citep[our data included 
the NUV data from $GALEX$ while][on the other hand, took the $24\ 
\rm{\mu m}$ data into consideration]{bon12}; our additional 
requirement of AGN contribution. Note that \cite{bon12} and our 
work both used the IMF of \cite{imf}, and the galaxy stellar masses 
from \cite{xue10} were also converted to those with the IMF of \cite{imf}. 
The M.A.D.\ we obtained should indicate this systematic/methodology 
uncertainty when estimating $M_{\ast}$ from SEDs. This uncertainty is 
much larger than the uncertainty introduced by the photometry error. 
Therefore, we adopted the normalized M.A.D.\ (NMAD), 
$\sigma_{\rm NMAD}=1.5\times \mathrm{M.A.D.} = 0.3$ dex, as the 
uncertainty of the galaxy total stellar mass \citep{mar06}. 

SFR estimation is subject to two major uncertainties: (1)\ the errors 
of the FIR fluxes; (2)\ the intrinsic uncertainty of using the K12 
template to estimate SFR is, as reported by K12, $0.17$ dex. We added 
these two uncertainties in quadrature to account for the full 
uncertainties of SFRs.

\section{Co-Evolution of AGNs and Their Host Galaxies}
\label{sec:result}
\subsection{Selection Biases}
\label{sec:bias}
We now discuss biases due to selection effects. First, our sources 
are selected based on AGN luminosity (as well as the luminosity 
contrast of the AGN and host galaxy in the optical/UV) and broad 
emission lines ($\rm{H}{\beta}$ or \MgII). Selecting galaxies around 
luminosity-limited AGNs leads to an Eddington bias in the 
$M_{\rm{BH}}-M_{\ast}$ relation, as detailed by \cite{lau07} (also 
see Section~\ref{sec:intro}). On the top of the \cite{lau07} 
bias, there is another bias on $M_{\rm BH}$ for our luminosity-limited 
AGN sample \citep{shk10}. Second, we required our sources be 
detected by \textit{Herschel}. The \textit{Herschel} detection limit 
effectively introduces a bias which acts in the opposite direction as 
the \cite{lau07} bias, since a SFR limit generally selects more massive 
host galaxies for an AGN with given $M_{\rm BH}$. 
\subsubsection{The Basic Model to Estimate Biases}
\begin{figure}
\plotone{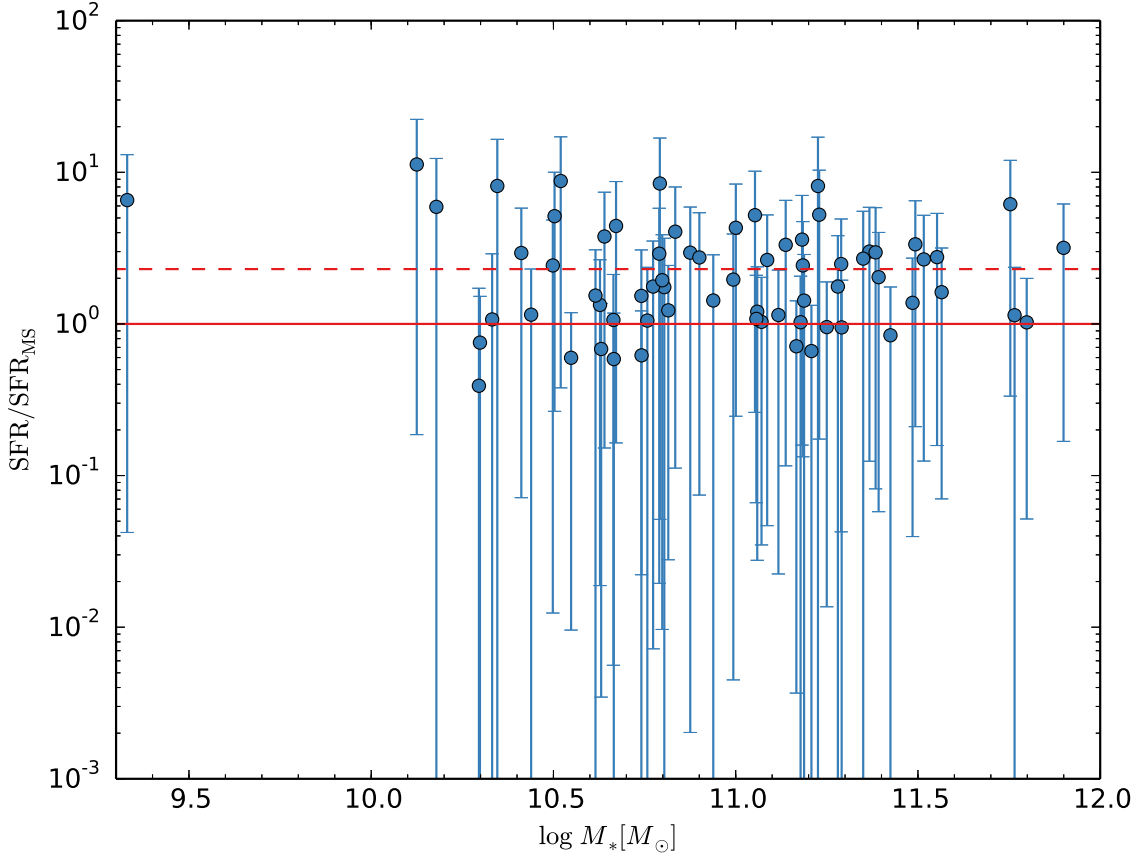}
\caption{
A comparison of SFR we measured with the expected SFR of star-forming 
main sequence galaxies with the same stellar mass. Our sources have a 
star formation activity which is marginally higher than that of the 
``main-sequence'' (the mean of $\mathrm{SFR/SFR_{ms}}$ is $2$). This 
offset is due to an Eddington bias which is driven by the \textit{Herschel} 
flux limit (see Section~\ref{sec:bias} for more details). The red dashed 
line represents the bias corrected $\mathrm{SFR/SFR_{ms}}$. 
}
\label{fig:sfrc}
\end{figure}

To quantify selection biases in our sample, we begin by assuming that 
the scatter in each quantity has a log-normal distribution \citep[as 
commonly assumed in, e.g.,][]{lau07, shk10}. According to the local 
SMBH mass-galaxy stellar mass relation $\langle m\rangle=c_1 s+c_2$, 
the distribution of the SMBH mass, $M_{\rm{BH}}\equiv10^m\ M_{\odot}$, 
at fixed galaxy stellar mass, $M_{\ast}\equiv10^s\ M_{\odot}$, is
\begin{equation}
\label{eq:gms}
p(m|s) = (2\pi \sigma_{\mu}^2)^{-0.5}\exp(-\frac{(m-(c_1 
s+c_2))^2}{2\sigma_{\mu}^2}) \\,
\end{equation}
where $\sigma_{\mu}$ is the intrinsic scatter. We also assume that 
our BLAGN host galaxies are drawn from the star-forming main sequence 
in which SFR correlates well with galaxy stellar mass. The motivations 
are as follows. First, we demonstrate below that this is a good assumption, 
as the apparent factor of two excess of SFR seen in Section~\ref{sec:sfr} 
and Figure~\ref{fig:sfrc} is fully consistent with the bias from the 
\textit{Herschel} sensitivity limit. Moreover, \cite{ros13b} also 
demonstrated that, after using simulations to model selection biases 
carefully, BLAGN hosts are consistent with being drawn from the 
star-forming main sequence relation. Note that BLAGNs are also 
observed to be a factor of $\gtrsim 5$ less common in quiescent galaxies 
than in blue star-forming hosts \citep{tru13,mat14}. Therefore, we can 
connect SFR with galaxy stellar mass by Equation~\ref{eq:ms}. The distribution 
of SFR ($\mathrm{SFR}\equiv10^{\lambda}\ M_{\odot}\ \rm{yr^{-1}}$) for 
fixed $s$ is 
\begin{equation}
\label{eq:msfr}
g(\lambda |s) = (2\pi \sigma_{F}^2)^{-0.5}\exp(-\frac{(\lambda-(a(s- 
10.5)+b))^2}{2\sigma_{F}^2}) \\,
\end{equation}
where $a=0.7-0.13z$, $b=0.38+1.14z-0.19z^2$ (see Equation~\ref{eq:ms}). 
The intrinsic scatter of the $\mathrm{SFR}-M_{\ast}$ relation 
$\sigma_{F}$ is roughly independent of both redshift and galaxy stellar 
mass \citep[e.g.,][]{whi12}. 

Equations~\ref{eq:gms} and \ref{eq:msfr} can be used to estimate the 
bias in the $M_{\rm{BH}}-M_{\ast}$ relation resulting from sample 
selections based on $M_{\rm{BH}}$ and SFR. While the SFR selection 
is directly related to the \textit{Herschel} sensitivity, the 
$M_{\rm{BH}}$ selection is a more complicated function of the AGN 
luminosity limit and the broad-line width limit (i.e., 
$v_{\rm FWHM}\geq 10^3\ \rm{km\ s^{-1}}$). Motivated by \cite{shg08}, 
\cite{tru11}, and \cite{lus12} we related $M_{\rm{BH}}$ and 
$L_{\rm Bol}$ by assuming our BLAGNs have a log-normal distribution 
of the Eddington ratio centered on $0.1$ with a scatter of 
$\sigma_{\rm L}$ dex and a cut off of $\leq 0.01$ and $\geq 1$.
 Using a different shape for the distribution (e.g., log-uniform) turns 
out to have little effect on our bias estimate. However, assuming a different 
mean Eddington ratio for the distribution does have a small effect; see 
Section~\ref{sec:mbias}. The FWHM of the emission line 
was related to $M_{\rm{BH}}$ and $L_{\rm Bol}$ with a log-Gaussian 
distribution centered on the value given by Equation~\ref{eq:mse} with 
a scatter of $\sigma_{\rm FWHM}$ \citep{shg08}. 
To mimic the $M_{\rm BH}$ bias, we calculated the virial 
$M_{\rm BH, vir}$ by using the distributions of $L_{\rm Bol}$ (converted 
to $\lambda L_{\lambda}$ assuming a bolometric correction of $5$) and 
$v_{\rm FWHM}$ and Equation~\ref{eq:mse}. Note that the scatter of the 
virial $M_{\rm BH}$ to true $M_{\rm BH}$ is $\sqrt{(0.5\sigma_{\rm L})^{2} 
+ (2\sigma_{\rm FWHM})^{2}} \simeq 0.4$ dex \citep{shk10}. We then 
apply the AGN luminosity and SFR cuts to the distribution of 
$M_{\rm BH, vir}/M_{\ast}$ to estimate the selection biases (i.e., a 
superposition of the Lauer et al. 2007 and the $M_{\rm{BH}}$ biases).

\begin{figure}
\plotone{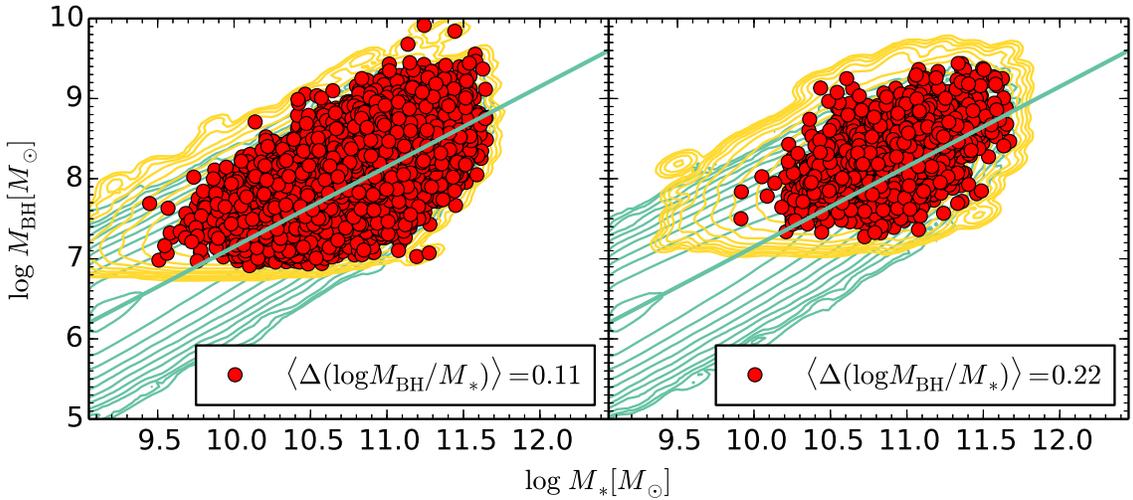}
\caption{
Monte Carlo simulation of biases in our data. Left (right) panel is for 
$z=1.0$ ($z=2.0$). The green and yellow contours correspond to the PDFs 
of our mock sample and sample A (which results from the mock sample with 
a AGN luminosity cut off). For sample B, which results from sample A with 
a SFR cut off (i.e., sample B and our data share the same selection bias.), 
the actual distribution of simulated sources in the $M_{\rm{BH}}-M_{\ast}$ 
plane is plotted. The solid green lines correspond to 
$\log M_{\rm{BH}}=\log M_{\ast}-2.85$. See Section~\ref{sec:bias} for more 
details.
}
\label{fig:bias}
\end{figure}

We performed Monte Carlo simulations (based on the previous arguments) 
to estimate the selection biases in our sample. We started by 
considering the galaxy stellar mass function of \cite{muz13} for 
star-forming galaxies:
\begin{equation}
\label{eq:smf}
\Phi(s)=(\ln 10)\Phi_{\ast}10^{(s-s_{\ast})(1+\alpha)}
\exp(-10^{s-s_{\ast}}) \\.
\end{equation}
$\alpha$, $s_{\ast}$, and $\Phi_{\ast}$ (at each redshift bin) can be 
found from Table 1 of \cite{muz13} (note that the absolute value of 
$\Phi_{\ast}$ is not important for the bias estimation). For each 
$M_{\ast}$ that was randomly drawn from the distribution of 
Equation~\ref{eq:smf}, $M_{\rm BH}$ was randomly generated from the 
probability density function (PDF) of Equation~\ref{eq:gms} with 
$c_1=1$, $c_2=-2.85$, and $\sigma_{\mu}=0.3$ \citep{hr04}. In addition, 
for each $M_{\ast}$, the SFR was randomly generated from the PDF of 
Equation~\ref{eq:msfr} with $\sigma_{F}=0.34$ \citep{whi12}. The AGN 
luminosity and line width were calculated as described in the previous 
paragraph by assuming $\sigma_{\rm L}=0.4$ and $\sigma_{\rm FWHM}=0.17$ 
(such that the uncertainty of the virial $M_{\rm BH}$ is $\sim 0.4$ 
dex; we also tested other values of $\sigma_{\rm L}$ and $\sigma_{\rm FWHM}$ 
and found that the bias is not sensitive to $\sigma_{\rm L}$ but 
depends more on the average value of the Eddington ratio 
distribution---see Section~\ref{sec:mbias} for details). 
We repeated the simulation $10^6$ times to get the simulated 
data, i.e., our mock sample. The PDF of these mock galaxies was 
estimated with kernel density estimation (green contour in 
Figure~\ref{fig:bias}). 

We applied our sample selections to this mock sample, starting 
with the limits on AGN luminosity and broad-line width. For AGNs 
in COSMOS, which form the bulk of our sample, these limits are 
$i_{\rm{AB}}<22.5$ and $v_{\rm{FWHM}}>10^3\ \rm{km s^{-1}}$ 
\citep{tru09a}. The resulting sub-sample is denoted as sample~A. 
With these flux and width limits, our sample A is strongly 
affected by the \cite{lau07} bias, as seen by the yellow contour 
in Figure~\ref{fig:bias}. In the left panel (for $z=1$), for 
example, we can clearly see a large offset between the 
$M_{\rm{BH}}-M_{\ast}$ relation of our sample A and that of our 
mock sample. More quantitatively, 
$\langle \Delta(\log M_{\rm{BH}}/M_{\ast})\rangle \equiv \langle 
\log(M_{\rm{BH}}/M_{\ast})\rangle - c_2 = 0.2$ for sample~A. 

Next, a redshift-dependent cut-off in SFR was also applied to sample 
A to mimic the \textit{Herschel} detection limit. The redshift-dependent 
SFR limit was estimated from the flux limit of \textit{Herschel}/PACS 
in COSMOS \citep[since most of our sources are from the COSMOS field; 
the limit is $\nu L_{\nu}(60\ \rm{\mu m})\simeq10^{45}\ \rm{erg\ 
s^{-1}}$ for $z=1$, see, e.g.,][]{ros12}. The resulting sub-sample 
is denoted as sample B and is plotted as red points in 
Figure~\ref{fig:bias}. For the left panel, the bias 
($\langle \Delta(\log M_{\rm{BH}}/M_{\ast})\rangle =0.11$) 
is much smaller compared with sample A. We also observed 
similar features in the right panel (for $z=2$): for sample A, 
$\langle \Delta(\log M_{\rm{BH}}/M_{\ast})\rangle = 0.3$; for 
sample B, $\langle \Delta(\log M_{\rm{BH}}/M_{\ast})\rangle =0.22$. 

We also compared the $\mathrm{SFR}-M_{\ast}$ relation of 
sample~B with the star-forming main sequence and found that 
sources in sample B show higher star formation activity. This 
result is simply due to selection effects: (1) there is intrinsic 
scatter in the ``main-sequence'' relation; (2) the galaxy stellar 
mass function is bottom heavy. That is, a star-forming galaxy with 
given SFR is more likely to be less massive, resulting in an 
apparent offset from the ``main-sequence'' relation. At $z\sim 1$, 
the median redshift for our sample, the 
$\langle \mathrm{SFR}/\mathrm{SFR}_{\rm ms}\rangle$ of sample B is 
$\sim 2$: the same as the mean offset we obtained in 
Section~\ref{sec:sfr} and Figure~\ref{fig:sfrc}. This result 
suggests that, after accounting for selection biases, our sources 
are consistent with being drawn from ``main-sequence'' galaxies. 
Therefore, it is valid to use the main-sequence relation in the 
simulations. 

For each redshift, we can perform similar simulations and therefore 
estimate the corresponding selection biases in sample B. Through this 
procedure, we estimated the selection biases of sample B as a function 
of redshift. We used this result to obtain the bias-corrected HR04 
relation (i.e., a summation of the HR04 relation and the bias). This 
bias-corrected relation will be included in Section~\ref{sec:msevol} 
(see the red solid line in Figure~\ref{fig:msr}). 

\subsubsection{More Detailed Consideration of Biases}
\label{sec:mbias}
The way we model the bias depends on the $M_{\rm{BH}}-M_{\ast}$ 
relation, the main-sequence relation, their intrinsic uncertainties, 
and the Eddington-ratio distribution. As will be shown in 
Section~\ref{sec:msevol}, our measured uncertainty in the 
$M_{\rm{BH}}-M_{\ast}$ relation is $\sim 0.3$ dex, consistent with 
our assumed value. The uncertainty of the star-forming ``main-sequence'' 
relation is not precisely known. If we instead use $\sigma_{F}=0.25$ as 
reported in \cite{whi12} for the ``normal'' star formation sequence, 
the bias of sample B at e.g., $z=1$ decreases to $\sim 0.06$ (the bias 
at $z\sim 2.0$ remains almost the same). We conservatively assume 
$\sigma_{F}=0.34$ when estimating the bias of our sample. The bias 
also depends on the connection between AGN activity and star formation. 
Let us assume, for example, the AGN luminosity is fully determined by 
galaxy properties (e.g., gas fueling) and is not directly linked with 
$M_{\rm BH}$. If so, there is no bias for the $M_{\rm{BH}}-M_{\ast}$ 
relation \citep[as already noted by][]{lau07}. The reason is simple: 
in this case, our sample would be based on galaxy properties rather 
than $M_{\rm BH}$, just like the local sample. Some works suggest a 
link between AGN activity and galaxy properties \citep[e.g,][and this 
work]{mul12,ros12,ros13a,ros13b,che13}. However, these AGN--galaxy 
relations may have large intrinsic scatter, perhaps making them 
inapplicable for modeling the bias in individual AGNs. 

We also have limited knowledge of the distribution of the Eddington ratio 
for BLAGNs \citep[e.g.,][]{kol06,shg08,tru09b,hop09,tru11,ksh13}. In our 
simulations, we also tried other Eddington-ratio distributions, e.g., 
a log-uniform distribution \citep[i.e., close to the ``fiducial'' model 
proposed by][]{hic14} or a log-normal distribution with a scatter 
of $0.8$ dex. We additionally required these distributions to be truncated 
at both $0.01$ and $1.0$. These two cut-offs are motivated by both the 
theory of accretion disks \citep[see, e.g,][]{adaf, yn14} and observations 
\citep[e.g.,][]{ho08,tru09b, tru11}. We found that all these 
distributions give similar results regarding selection biases as 
long as the mean value of these distributions is the 
same (i.e., $0.1$). If we instead assume a smaller mean value for 
the Eddington ratio, the selection biases increase (and vice versa if we 
assume a larger mean value for Eddington ratio). 
This is simply due to the fact that the selection limit of $M_{\rm BH}$ 
increases with decreasing Eddington ratio and the slope of the 
$M_{\rm BH}$ function becomes steeper with increasing $M_{\rm BH}$. 
To illustrate its effect, we adopted a power-law distribution of 
the Eddington ratio with a slope of $-1.0$ \citep{bon12} and a cut-off 
of $>0.01$ and $<1.0$. This distribution has a mean Eddington ratio 
of $\sim 0.02$. The resulting selection bias on the $M_{\rm BH}-M_{\ast}$ 
relation is $0.22$ ($0.32$) at $z=1$ ($z=2$). 
Future constraints on the Eddington ratio distribution are crucial for a more 
reliable estimation of the biases.  Regardless, we emphasize that our conclusions 
in Section~\ref{sec:msevol} would not be significantly changed if the 
selection bias is essentially zero or as large as $\sim 0.3$. 

The detectability of a BLAGN actually depends on the luminosity contrast 
between the AGN and the host galaxy, since this determines whether we 
can detect the broad H$\beta$ and/or \MgII\ emission lines (and therefore 
measure $M_{\rm{BH}}$). This effect has not been discussed in previous 
simulations of the \cite{lau07} bias. To check the luminosity contrast 
in our simulated samples, we used the 3D-HST catalogs 
\citep{3dhst1, 3dhst2}.\footnote{\url{http://3dhst.research.yale.edu/Home.html}} 
The 3D-HST catalogs have redshift, stellar mass, age, SFR, and rest-frame 
colors for inactive galaxies and can be used to assign galaxy properties 
to our simulated samples. For each redshift bin, we started by dividing 
our mock sample into galaxy stellar mass bins. As a second step, for each 
redshift and galaxy stellar mass bin, we selected main-sequence galaxies 
(defined by Equation~\ref{eq:ms} and its $3\sigma$ uncertainty) from the 
3D-HST catalogs whose redshifts and galaxy stellar masses lie within 
the bin. Then we used the selected sources to determine the distributions 
of the galaxy luminosity at (rest-frame) $2800\ \rm{\AA}$ 
and the SDSS--$g$ band (whose effective wavelength is $4718.8\ \rm{\AA}$, 
close to the wavelength of H$\beta$) for each bin. Finally, we 
assigned the (rest-frame) galaxy luminosity to our mock samples according 
to the distributions. We again applied the limits in AGN luminosity, 
FWHM, and SFR. For the resulting samples, we calculated the distribution 
of the luminosity ratio of the AGN to the host galaxy at $2800\ \rm{\AA}$ 
(for sources at $z>0.88$, whose $M_{\rm BH}$ are measured using \MgII) 
or the SDSS-$g$ band (for sources at $z<0.88$ whose $M_{\rm BH}$ are 
measured using H$\beta$). We found that for the mock sample A, there 
is a fraction of sources with $L_{\rm{AGN}}/L_{\rm{Gal}}<0.1$ at these 
wavelengths. However, no source in the mock sample B (i.e., the one which 
suffers the same selection biases as our data) 
has a luminosity ratio $<0.1$. This suggests that our assumption of the 
AGN contribution to the total SED in Section~\ref{sec:mgl} is appropriate 
and does not introduce additional selection biases to our results. 

In order to measure the galaxy total stellar mass via two-component 
SED fitting (see Section~\ref{sec:mgl}), the galaxy must be detectable 
and not outshone by the AGN in the $K$ band. This requirement will 
introduce an upper limit to $M_{\rm{BH}}/M_{\ast}$. We calculated this 
limit for the galaxy emission to be at least $10\%$ of the AGN emission 
in the $K$ band. Assuming an AGN bolometric correction of $\sim 10$ at 
the $K$ band \citep{R06} and an Eddington ratio of $0.1$ \citep[see 
Figure~\ref{fig:bolmbh} and e.g.,][]{kol06, tru11, lus12}, the $K$-band 
AGN luminosity is
\begin{equation}
L^{\rm{K}}_{\rm{AGN}} = 1.3\times 10^{36}M_{\rm{BH}} \\.
\end{equation}
Meanwhile, the $K$-band galaxy luminosity can be written as 
\begin{equation}
L^{\rm{K}}_{\rm{Gal}} = \frac{M_{\rm{Gal}}}{\gamma}L_{\odot} \\,
\end{equation}
where $\gamma$ is the mass-to-light ratio. We adopted $\log(\gamma) 
\sim -0.6$ for our $z\sim 1$ star-forming galaxies \citep[][with a 
correction for the IMF of Chabrier 2003]{arn07}. Note that 
there is a weak dependence of $\gamma$ on redshift, although this 
is not significant over the redshift range of our sources. With 
these parameters the requirement of 
$L^{\rm{K}}_{\rm{Gal}}>0.1 L^{\rm{K}}_{\rm{AGN}}$ indicates that 
$M_{\rm{BH}}/M_{\ast}$ cannot exceed $0.16$. This 
$M_{\rm{BH}}/M_{\ast}$ limit is more than two orders of magnitude 
larger than the mean value of the $M_{\rm{BH}}-M_{\ast}$ relation 
(which suggests $\langle M_{\rm{BH}}/M_{\ast}\rangle \sim 0.001$). 
Therefore, we expect this limit should not introduce significant 
biases to our data. We also directly test the effect of the 
$L^{\rm{K}}_{\rm{Gal}}>0.1 L^{\rm{K}}_{\rm{AGN}}$ limit by assigning 
the rest-frame $K$-band luminosity to our Monte Carlo simulations 
of the $M_{\rm{BH}}-M_{\ast}$ relation, again finding that the galaxy 
detection limit does not introduce significant bias.

\subsection{Black Hole Mass-Galaxy Total Stellar Mass Relation of 
Star-forming Galaxies}
\label{sec:msevol}
\begin{figure}
\plotone{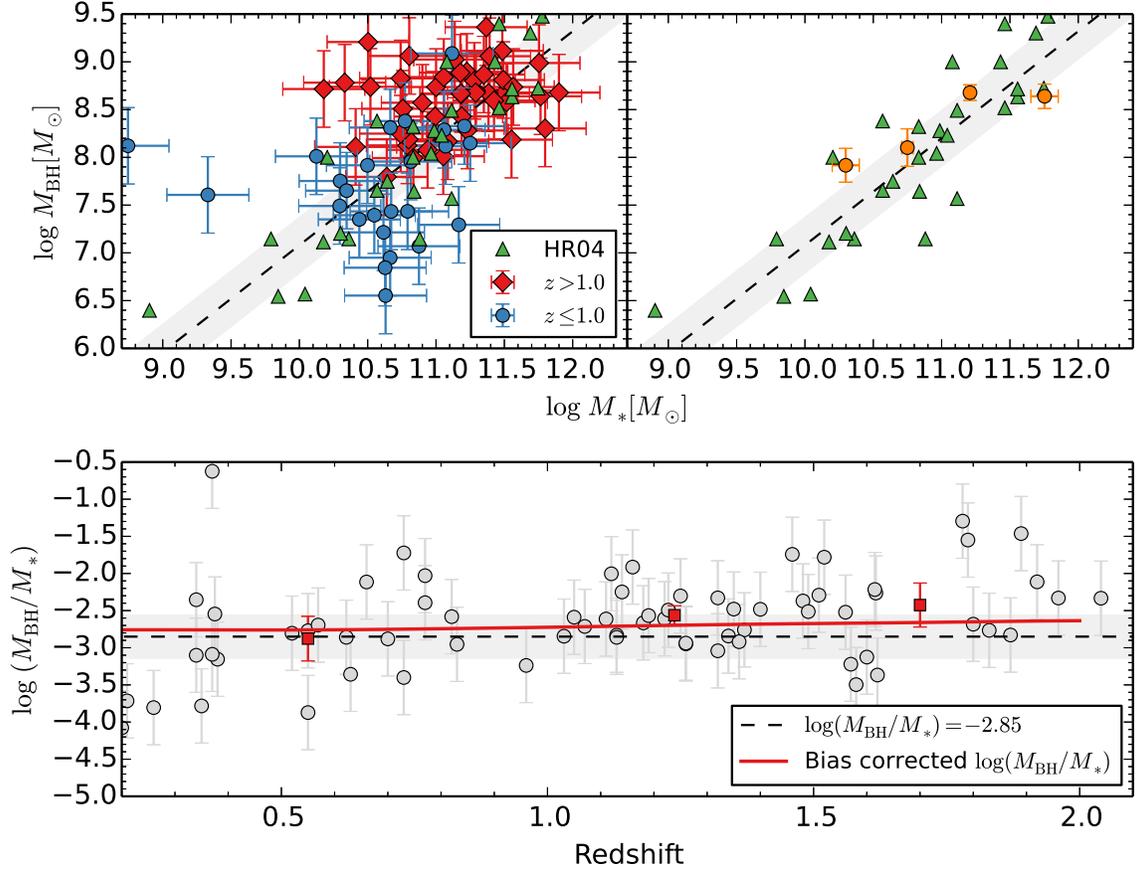}
\caption{
Upper-left panel: The distribution of our BLAGNs in the 
$M_{\rm{BH}}-M_{\ast}$ plane. 
Upper-right panel: Sources are binned by $M_{\ast}$; four orange 
points represent the median values of $M_{\rm BH}$ and $M_{\ast}$ and 
their $1\sigma$ uncertainties (estimated via bootstrapping) for sources 
within four bins: $M_{\ast}<10^{10.5}\ M_{\odot}$, 
$10^{10.5}\ M_{\odot}\leq M_{\ast}< 10^{11}\ M_{\odot}$, 
$10^{11}\ M_{\odot}\leq M_{\ast}< 10^{11.5}\ M_{\odot}$, and 
$M_{\ast}\geq10^{11.5}\ M_{\odot}$. The triangles in both panels 
represent sources from HR04. 
Lower panel: Logarithm of the ratio of the SMBH mass to the galaxy total 
stellar mass vs.\ redshift. Three red squares represent the mean value 
of the mass ratio and its $2\sigma$ uncertainty for sources with $z<1.0$, 
$1.0\leq z<1.5$, and $z\geq 1.5$. The red solid line represents the local 
HR04 relation as it would be observed under the biases of our sample. 
The lines and shaded regions in all panels represent the local SMBH 
mass-bulge mass relation by HR04 and its uncertainty. Our sources are 
indeed consistent with the biased HR04 relation, indicating no 
significant evolution of the $M_{\rm{BH}}-M_{\ast}$ relation. 
}
\label{fig:msr}
\end{figure}

In this subsection, we explore the $M_{\rm{BH}}-M_{\ast}$ relation 
of BLAGNs in star-forming galaxies. The upper-left panel of 
Figure~\ref{fig:msr} plots the distribution of our sources in the 
SMBH mass--galaxy total stellar mass plane. For comparison, we 
also include the sample of local inactive galaxies from HR04 and 
the local SMBH mass-bulge mass relation of HR04\footnote{\cite{kor13} 
point out that the HR04 SMBH-to-bulge mass ratio is probably 
underestimated by a factor of $\sim 2$ (see Section~$6.6$ in their 
paper). However, our single 
epoch virial SMBH mass estimators are calibrated by the local 
$M_{\rm{BH}}-\sigma$ relation of \cite{tre02} which is consistent 
with the HR04 SMBH mass-bulge mass relation. That is, the geometry 
factor which converts the virial product into a SMBH mass is derived 
from a $M_{\rm{BH}}-\sigma$ relation that is consistent with the HR04 
relation \citep{onk04}. Therefore we retain the 
older HR04 relation for consistency with the single epoch virial 
SMBH mass estimates. Updating to the \cite{kor13} relation would 
require similar updating single epoch virial SMBH estimates by 
(roughly) the same factor, resulting in no difference to our 
conclusions regarding evolution.} and its $1\sigma$ uncertainty. 
As seen from the upper-left panel of Figure~\ref{fig:msr}, our 
sample is consistent with the HR04 relation. In fact, only 
$13/69$ sources lie outside the $1\sigma$ uncertainty of the HR04 
relation when considering error bars. A Mann-Whitney U test of 
$M_{\rm{BH}}/M_{\ast}$ on our sample and the local inactive 
galaxies also indicates that the two samples are statistically 
consistent (with a null probability $p=0.2$). The same conclusion 
holds for our low-redshift sub-sample, although considering the 
high-redshift subsample alone results in a small (but statistically 
significant) difference from the HR04 relation. However, this is 
likely due to selection bias since the bias increases with redshift. 
Indeed, if we take the selection bias (see Section~\ref{sec:bias}) 
into consideration, the Mann-Whitney U test indicates that there is no 
significant difference between the high-redshift sub-sample and 
the local sample of HR04 (with $p=0.5$). We also binned our sources 
by $M_{\ast}$. Four bins were created: $M_{\ast}<10^{10.5}\ M_{\odot}$, 
$10^{10.5}\ M_{\odot}\leq M_{\ast}< 10^{11}\ M_{\odot}$, 
$10^{11}\ M_{\odot}\leq M_{\ast}< 10^{11.5}\ M_{\odot}$, and 
$M_{\ast}\geq10^{11.5}\ M_{\odot}$. We calculated the median values 
of $M_{\rm{BH}}$ and $M_{\ast}$ and their $1\sigma$ uncertainties 
for each bin. The upper-right panel of Figure~\ref{fig:msr} plots 
the results. As a reference, we again include the sample of 
local inactive galaxies from HR04 and the local SMBH mass-bulge 
mass relation of HR04 and its $1\sigma$ uncertainty. As seen 
from this panel, the binned data are consistent with the local 
SMBH mass-bulge mass relation of HR04. 

We can also quantify possible evolution by testing the correlation 
between $M_{\rm{BH}}/M_{\ast}$ and redshift. The lower panel of 
Figure~\ref{fig:msr} plots this mass ratio and its uncertainty as 
a function of redshift. As a reference, we also plot the local mass 
ratio (black dashed line) and its $1\sigma$ uncertainty. 
For each redshift, we estimated the corresponding selection bias in 
Section~\ref{sec:bias}. The red solid line in the lower panel of 
Figure~\ref{fig:msr} shows the expected HR04 relation after considering 
the selection bias. We also calculated the mean value of 
$M_{\rm{BH}}/M_{\ast}$ and its $2\sigma$ uncertainty for sources with 
$z<1$, $1\leq z<1.5$, and $z> 1.5$; red squares show the result. 
Comparing the red squares with the red solid line, we find our sample 
shares the same mean value of $M_{\rm{BH}}/M_{\ast}$ with the 
bias-corrected local one. 
 
We also used the simulation-extrapolation (SIMEX) technique \citep{simex} 
to estimate the intrinsic scatter (subtracting the measurement errors) of 
the distribution of $M_{\rm{BH}}/M_{\ast}$. The SIMEX estimation gives a 
scatter of $0.36$ dex for our sample. This indicates that the intrinsic 
scatter of $M_{\rm{BH}}/M_{\ast}$ for our sample is not significantly 
larger than the local value. Our observed scatter in 
$M_{\rm{BH}}/M_{\ast}$ is somewhat affected by the biases, so we do not 
put too much emphasis on the similarity in scatter. Still, the similar 
scatter suggests that our estimation of selection bias is reliable, and 
that our sample is consistent with the HR04 relation (also see 
Section~\ref{sec:dis1}). This non-evolution result agrees well with, 
e.g., \cite{jah09, sch13}, but it contrasts with some other works, 
e.g., \cite{mer10}. The latter claimed an evolution of 
$M_{\rm{BH}}/M_{\rm{\ast}}$ at the $>5\sigma$ significance level, 
although this is probably only due to the selection biases 
\citep{mer10, sw14}.

\section{Time Evolution of the SMBH Mass--Galaxy Total Stellar Mass 
Relation}
\label{sec:tmevol}
In the previous section, we find that the $M_{\rm{BH}}-M_{\ast}$ 
relation does not evolve with $z$ up to $z\sim 2$. In addition, 
the X-ray and FIR data allow us to calculate the current \textit{mass 
growth rate} of both AGNs and their host galaxies. We use these mass 
growth rates to assess if our galaxies and AGNs will remain on 
or move away from the established $M_{\rm{BH}}-M_{\ast}$ 
relation.

\begin{figure}
\plotone{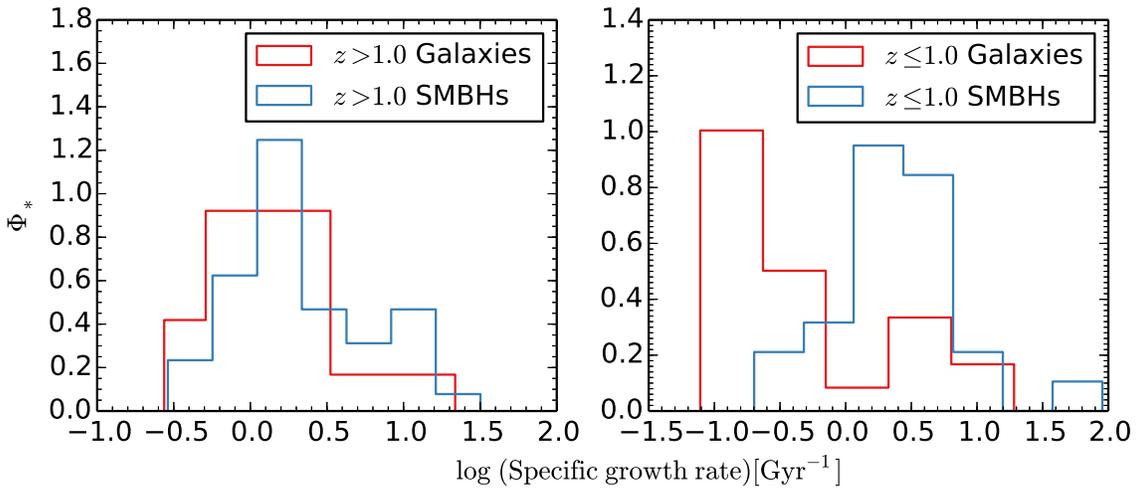}
\caption{
Distributions of the specific SMBH and galaxy mass growth rates. The left 
(right) panel is for the high-(low-)redshift sub-sample. The specific galaxy 
growth rate distribution evolves significantly as a function of redshift 
while the specific SMBH growth rate distribution does not change dramatically. 
Note that $\Phi_{\ast}$ is the probability density function (i.e., 
$\int \Phi_{\ast}(x) \, \mathrm{d}x=1$). 
}
\label{fig:gwr}
\end{figure}

We start by exploring the distribution of the specific SMBH mass growth 
rate (defined as $\mathrm{s}\dot{M}\equiv\dot{M}/M_{\rm{BH}}$) and the 
specific galaxy total stellar mass growth rate (defined as $\mathrm{sSFR} 
\equiv \mathrm{SFR}/M_{\ast}$). Figure~\ref{fig:gwr} plots the specific 
mass growth rate distributions for the high-redshift and low-redshift 
sub-samples. There are two points to emphasize: (1)\ in both sub-samples, 
the instantaneous specific SMBH growth rate is not smaller than that 
of the galaxy (especially for the low-redshift sub-sample, the instantaneous 
specific growth rate of the SMBH is much larger than that of the host galaxy); 
(2)\ the specific SMBH growth rate does not evolve much (the mean offset is 
$\sim 0.2$ dex) while the specific galaxy growth rate decreases significantly 
(the mean offset is $\sim 0.8$ dex) from high to low redshift. 
That is, our results indicate that the apparent $\mathrm{s}\dot{M}/\mathrm{sSFR}$ 
seems to increase from high to low redshift. This turns out to be largely due to 
selection effects; after modeling the biases in Section~\ref{sec:dis}, we find that 
the data are consistent with a uniform AGN duty cycle at both high and low 
redshift. 

In addition, for both sub-samples, the specific black hole mass growth rate 
distributions show negligible tails at the growth rate 
$<0.2(10^{-0.7})\ \rm{Gyr^{-1}}\simeq 0.01\dot{M}_{\rm{Edd}}$ (see 
Figure~\ref{fig:gwr}), where $\dot{M}_{\rm{Edd}}=L_{\rm{Edd}}/0.1c^2$. These 
negligible tails agree with the conclusion of, e.g., \cite{col06, kol06, tru11, 
lus12} that BLAGNs have a minimum Eddington ratio of $\sim 0.01$. 

\begin{figure}
\plotone{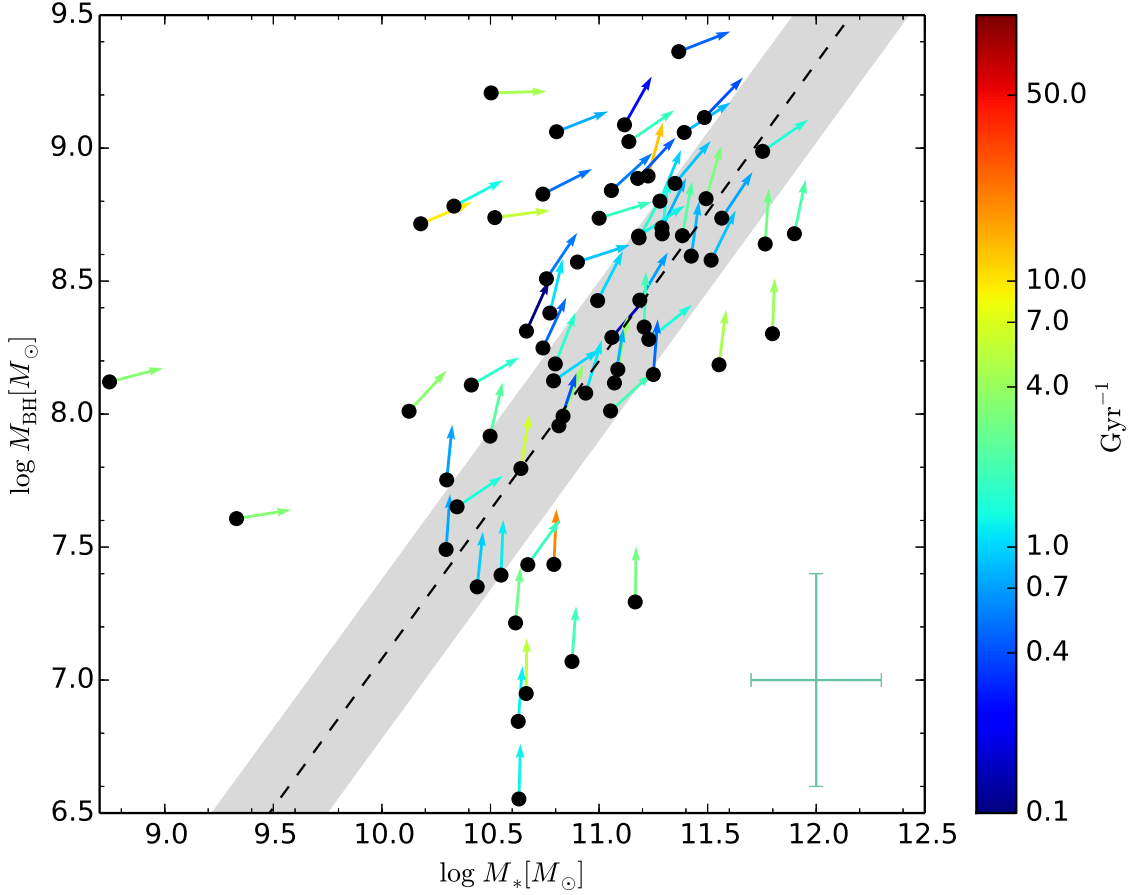}
\caption{
The ``flow patterns'' of SMBHs and their host galaxies in the SMBH mass-galaxy 
total stellar mass plane. Arrows represent the direction of the evolution (45 
degrees indicates $\log (\mathrm{s}\dot{M}/\mathrm{sSFR})=0$). Colors indicate 
the absolute value of the total specific growth rate (i.e., 
$\sqrt{\mathrm{s}\dot{M}^2+\mathrm{sSFR}^2}$) which spans $>2$ orders of 
magnitude. The dashed line and shaded area represent the local SMBH mass-bulge 
mass relation of HR04 and its uncertainty. The cross indicates the $1\sigma$ 
uncertainties of $M_{\rm{BH}}$ and $M_{\ast}$. The length of the vectors are 
arbitrary, since the ``flow patterns'' over a long timescale ($\simeq\ 1\ \mathrm{Gyr}$) 
depends on AGN duty cycle (see Section~\ref{sec:dis3}). 
}
\label{fig:fpa}
\end{figure}

\begin{figure}
\plotone{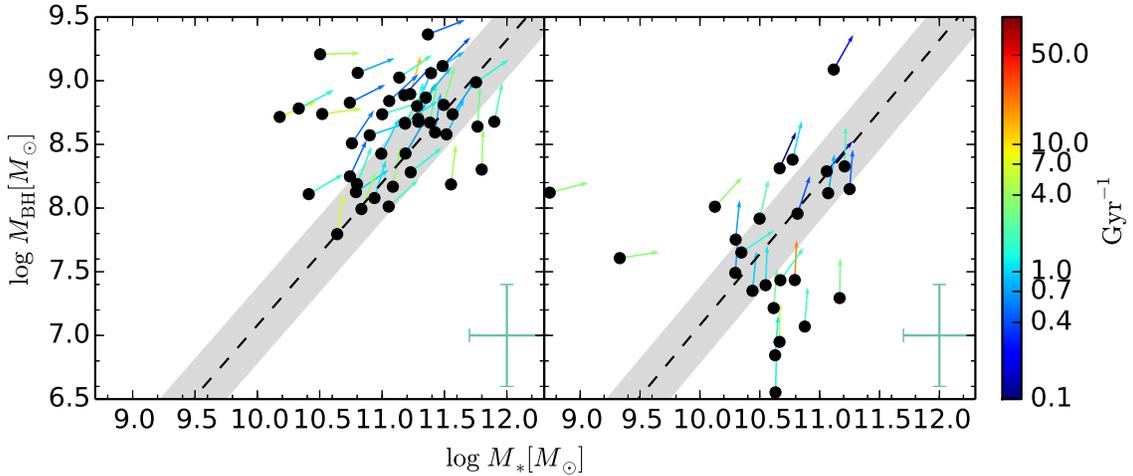}
\caption{
The same as Figure~\ref{fig:fpa}, but for the high-redshift (left panel) 
and low-redshift (right panel) sub-samples.
}
\label{fig:fps}
\end{figure}

We use these specific growth rates to investigate the ``flow patterns'' 
\citep[following][]{mer10} of our sample in the $M_{\rm{BH}}-M_{\ast}$ 
plane. Figures~\ref{fig:fpa} and \ref{fig:fps} plot the results. The 
direction of the vector is defined by $\mathrm{s}\dot{M}/\mathrm{sSFR}$ 
while the color map indicates the absolute specific growth rate 
(i.e., a summation of the specific SMBH mass growth rate and the specific 
galaxy total stellar mass growth rate in quadrature) along the vector 
(which spans more than two orders of magnitude). We again include the 
local HR04 relation and its $1\sigma$ uncertainty as the dashed line and 
the shaded region. For simplicity, these figures do not show the 
uncertainties of $\mathrm{s}\dot{M}/\mathrm{sSFR}$ on the plot (the 
uncertainties can instead be seen in Figure~\ref{fig:grms}). In general, 
AGNs and galaxies which are outliers in $M_{\rm{BH}}/M_{\ast}$ tend to 
have evolutionary vectors whose directions are anti-correlated 
with their positions: galaxies with over-massive SMBHs tend to have lower 
$\mathrm{s}\dot{M}/\mathrm{sSFR}$, and galaxies with under-massive 
SMBHs tend to have higher $\mathrm{s}\dot{M}/\mathrm{sSFR}$. This 
anti-correlation behavior can be seen not only for the whole sample 
(Figure~\ref{fig:fpa}) but also for each sub-sample (the left and right panels 
of Figure~\ref{fig:fps}). As will be shown in Section~\ref{sec:dis2}, this result 
is not solely due to selection effects. In Sections~\ref{sec:dis2} and 
\ref{sec:dis3}, we demonstrate the anti-correlation between the (instantaneous) 
growth rate and mass ratios, if extended over some time with an AGN duty 
cycle of $0.1$, implies that AGN activity and star formation could maintain 
the $M_{\rm BH}-M_{\ast}$ relation.

\section{Discussion}
\label{sec:dis}
\subsection{Building the Local $M_{\rm{BH}}-M_{\rm{Bul}}$ Relation}
\label{sec:dis1}
In Section~\ref{sec:msevol} we showed that our sample exhibits no apparent 
redshift evolution in the $M_{\rm{BH}}-M_{\ast}$ relation. Therefore, our 
results along with, e.g., \cite{jah09} and \cite{sch13} suggest that the 
$M_{\rm{BH}}-M_{\ast}$ relation does not evolve strongly with redshift. 
As stated before, the local inactive galaxies of HR04 are bulge-dominated 
($M_{\rm{Bul}}\simeq M_{\ast}$). Meanwhile, our sources (at least a 
significant fraction) are likely to have substantial disk components. 
First, our sources show strong star formation activity which indicates 
non-negligible disk components (for example, Lang et al. 2014 suggest that 
star-forming galaxies are likely to have smaller bulge-to-total ratios). In 
addition, for the X-ray luminosity and redshift ranges of our sources, 
a substantial fraction of AGNs reside in disk galaxies \citep[i.e., 
$M_{\rm{Bul}}\ll M_{\ast}$; see, e.g.,][]{gab09, geo09, cis11, koc12}. 
Furthermore, we have also cross-matched our sample with that of 
\cite{gab09} (for COSMOS sources) and \cite{sch11} (for CDF-S sources) 
and found that five out of the fourteen matched sources have 
S\'{e}rsic index much smaller than $2.5$. Therefore, the 
$M_{\rm{BH}}-M_{\rm{Bul}}$ relation is likely to evolve (at least mildly) 
with redshift. As suggested by \cite{jah09}, a redistribution of stellar 
mass from disk to bulge is required to build the local 
$M_{\rm{BH}}-M_{\rm{Bul}}$ relation. 

Our results suggest that in the next $\sim 10\ \rm{Gyr}$ a typical 
galaxy in our sample would (1)\ increase its galaxy total stellar 
and SMBH masses accordingly; (2)\ transfer much of its stellar mass 
from its disk to its bulge. The former indicates that gas fueling 
controls both AGN activity and star formation (see Section~\ref{sec:dis2} 
for more details). For the redistribution of stellar mass from disk to 
bulge, possible mechanisms are, e.g., violent disk instabilities 
\citep[which are driven by cold gas, see e.g.,][]{dek09} and/or galaxy 
mergers \citep{hop10}. The former mechanism is proposed since our sources 
are likely to have substantial cold gas \citep{ros13a, vit14}, although 
observational evidence of violent disk instabilities in AGN hosts is mixed 
\citep{bou12, tru14}. The latter mechanism is also attractive as it can 
not only redistribute stellar mass from disk into bulge but also help 
reduce the intrinsic scatter in the $M_{\rm{BH}}-M_{\ast}$ relation 
\citep{jah11}. Note that our AGNs are more likely fueled by cold gas 
rather than triggered by galaxy-galaxy mergers \citep[see, e.g.,][and 
our next Section]{gab09, geo09, cis11, koc12}. Therefore, if galaxy-galaxy 
mergers do play a role, they are likely to be dry mergers rather than 
gas-rich wet mergers. 

Currently, our estimation (see Section~\ref{sec:msevol}) indicates 
that the intrinsic scatter of the $M_{\rm{BH}}-M_{\ast}$ relation 
is not significantly larger than that of the HR04 relation. However, 
this estimation is also subject to selection biases. Using our 
simulated samples in Section~\ref{sec:bias}, we found that a range 
of intrinsic scatter (e.g., $0.36-0.5$ dex) can reproduce the 
observed scatter. On the other hand, our selection corrections in 
Section~\ref{sec:msevol} are made by assuming the $M_{\rm{BH}}-M_{\ast}$ 
relation has the same intrinsic scatter as the HR04 relation and our 
data are consistent with the bias-corrected HR04 relation. This 
indicates that the intrinsic scatter of our sample should not be 
significantly larger than that of the HR04 relation (i.e., the 
intrinsic scatter is at most $0.5$ dex). An average of one major 
merger for each of our galaxies is enough to reduce the intrinsic 
scatter by a factor of $\sqrt{2}$ \citep{jah11} if the intrinsic 
scatter is indeed $0.5$ dex rather than $\sim 0.3$ dex. 

\subsection{Growth of SMBHs and Host Galaxies}
\label{sec:dis2}
\begin{figure}
\plotone{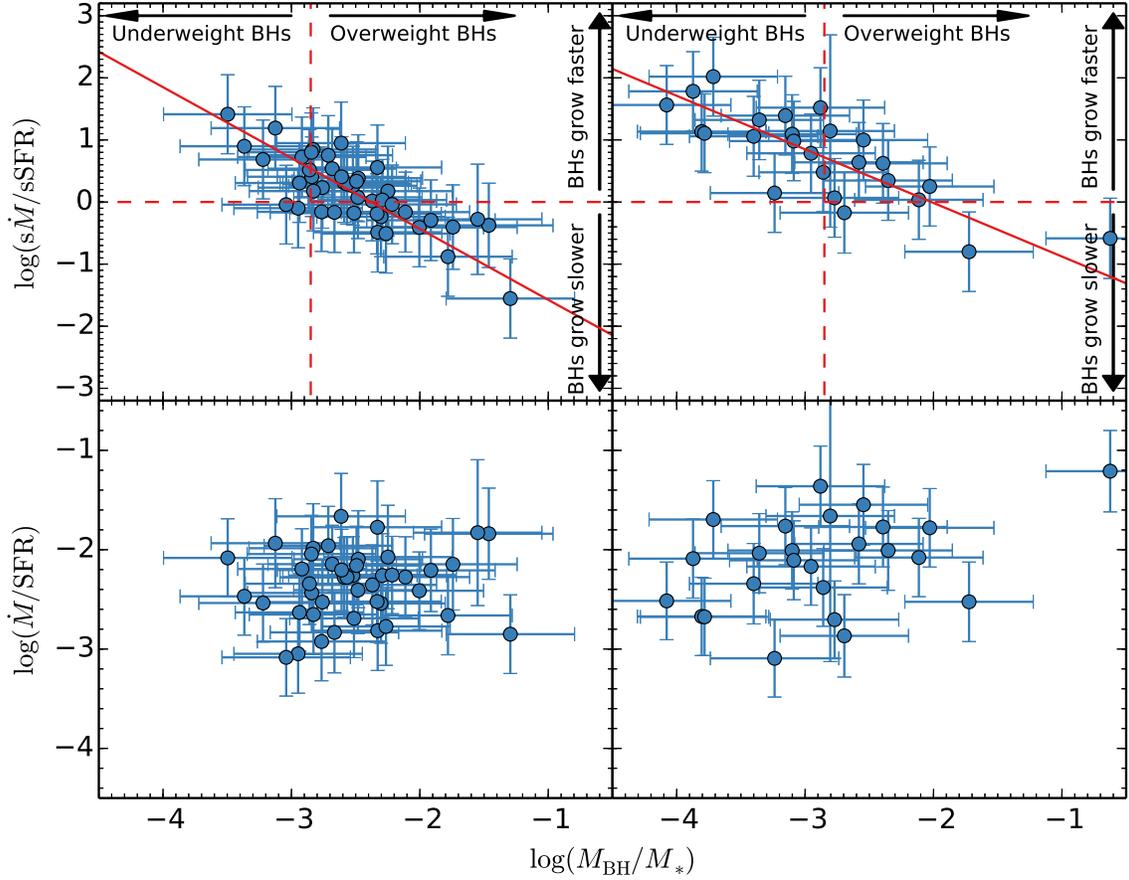}
\caption{
Upper panels: The ratio of the SMBH to the host-galaxy specific mass growth rate 
vs.\ the ratio of the SMBH mass to the host-galaxy total stellar mass. Left (right) 
panels are for the high-(low-) redshift sub-samples. The red solid line in each 
panel represents the best fit. The best fit indicates an anti-correlation between 
$M_{\rm{BH}}/M_{\ast}$ and $\mathrm{s}\dot{M}/\mathrm{sSFR}$ is significant 
(null probability $p<10^{-6}$).  The red vertical dashed line in each panel 
represents the local mass ratio from HR04. The red horizontal dashed line in 
each panel is an indicator of $\mathrm{s}\dot{M}/\mathrm{sSFR}=1$, i.e., the 
specific SMBH accretion rate and the specific SFR are equal. Note that there are 
more sources above the horizontal line. This asymmetric distribution indicates 
that, if the growth rates of our sources persist over some length of 
time, AGN duty cycle is less than unity (i.e., the lifetime of an active SMBH is 
smaller than the lifetime of star formation), otherwise the SMBH will become 
over-massive as we evolve our sources. 
Lower panels: The ratio of the SMBH accretion rate to SFR vs.\ the ratio of 
the SMBH mass to the host-galaxy total stellar mass. Again, left (right) panels 
are for the high-(low-) redshift sub-samples. A linear fit indicates 
that $\dot{M}/\mathrm{SFR}$ is a constant (with scatter) over $M_{\rm{BH}}/M_{\ast}$.
}
\label{fig:grms}
\end{figure}

In Section~\ref{sec:tmevol}, we demonstrated that AGNs which are offset 
from the $M_{\rm{BH}}-M_{\ast}$ relation tend to have (instantaneous) 
evolutionary vectors which are anti-correlated with their mass offsets. 
That is, galaxies with over-massive (under-massive) SMBHs tend to have 
lower (higher) $\mathrm{s}\dot{M}/\mathrm{sSFR}$. We quantify the 
anti-correlation between $M_{\rm{BH}}/M_{\ast}$ and 
$\mathrm{s}\dot{M}/\mathrm{sSFR}$ by performing a linear fit (via 
chi-squared minimization, taking the uncertainties into consideration). 
Results are plotted in the upper panels of Figure~\ref{fig:grms}. The red 
vertical dashed lines represent the local $M_{\rm{BH}}/M_{\ast}$. The 
red horizontal dashed lines correspond to 
$\mathrm{s}\dot{M}/\mathrm{sSFR}=1$. There are more sources above 
the horizontal lines. If the growth rate rates persist over some 
length of time, then this asymmetric distribution indicates the AGN duty 
cycle is less than unity (i.e., the lifetime of an active SMBH is 
smaller than the lifetime of star formation); otherwise the SMBH will 
become over-massive as we evolve our sources. For example, if we 
assume our sources keep their current $\dot{M}$ and SFR for some 
timescales $t_{\rm AGN}$ and $t_{\rm SF}$, we expect BLAGN hosts to 
evolve as 
$\Delta(\log M_{\rm BH}/M_{\ast})=
\log(\frac{1+\mathrm{s}\dot{M}t_{\rm AGN}}{1+\mathrm{sSFR}t_{\rm SF}})$. 
To maintain the $M_{\rm{BH}}-M_{\ast}$ relation, we expect that, on average, 
$t_{\rm AGN}<t_{\rm SF}$. In Section~\ref{sec:dis3}, we derive the appropriate 
AGN duty cycle ($t_{\rm AGN}/t_{\rm SF}$) such that, under appropriate 
assumptions about the AGN activity and star formation histories, the 
$M_{\rm{BH}}-M_{\ast}$ relation is maintained. The red solid lines are for 
the best fits of the anti-correlation: for the high-redshift sub-sample, the best 
fit is $y = -2.7(\pm 0.3) - 1.14(\pm 0.12)x$; for the low-redshift sub-sample, 
the best fit is $y = -1.7(\pm 0.4) - 0.86(\pm 0.12)x$. In each sub-sample, 
the slopes are consistent with $-1$ and differ from zero at a high 
significance level (for the high-redshift sub-sample, the null probability is 
$p=4\times 10^{-12}$; for the low-redshift one, $p=2\times 10^{-7}$). 

The anti-correlation between the \textit{specific} growth rate ratio 
(s$\dot{M}$/sSFR) and the mass ratio, with slope of $-1$, implies that the 
\textit{absolute} growth rate ratio ($\dot{M}$/SFR) is not correlated with the 
mass ratio and is a constant (with small scatter).  We demonstrate 
this in the lower panels of Figure~\ref{fig:grms}. A linear fit (via chi-squared 
minimization) between $\dot{M}/\mathrm{SFR}$ and $M_{\rm{BH}}/M_{\ast}$ 
results in a slope of $0.27\pm 0.3$, i.e., statistically consistent with 
zero. The relatively narrow distribution of the $\dot{M}$/SFR ratio is 
in agreement with \citet{mul12}, after accounting for the AGN duty cycle (see 
Section~\ref{sec:dis3}). 

To test whether this anti-correlation and the narrow 
distribution of $\dot{M}$/SFR ratio are simply due to 
selection effects we again turn to the Monte Carlo simulations in 
Section~\ref{sec:result}. These simulations created mock samples 
with $M_{\rm{BH}}$, $M_{\ast}$, $\dot{M}$, and SFR following the 
same selection criteria as our data. We tested three Eddington-ratio 
distributions which are log-normal with scatter ($\sigma_{\rm L}$) 
of $0.8$, $0.6$ and $0.4$, respectively. Meanwhile, we use the same 
$\mathrm{SFR}-M_{\ast}$ and $M_{\rm BH}-M_{\ast}$ relations and 
scatter as in Section 4, and $\sigma_{\rm FWHM}$ is set such that 
the uncertainty of the virial $M_{\rm BH}$ is $\simeq 0.4$ dex. We 
created 69 mock samples, one at each redshift of the 69 observed 
BLAGNs, with $4\times 10^5$ sources in each mock sample. 

The initial mock sample (i.e., the one without any selection cut-offs) 
at each redshift does not show an anti-correlation between the 
ratio of the masses and their growth rates. Indeed, Spearman's 
correlation test indicates a coefficient of $\rho \sim -0.002$ for 
$\sigma_{\rm L}=0.4$ or weaker depending on $\sigma_{\rm L}$). 
We then applied the same (redshift-dependent) cut-offs presented 
in Section~\ref{sec:bias} to these mock samples. Finally, we randomly 
selected $69$ sources from the 69 mock samples (i.e., selected one 
source per mock sample) and calculated the corresponding $\rho$ (for 
our BLAGNs sample, $\rho=-0.78$) and the scatter of the 
distribution of $\dot{M}/\mathrm{SFR}$ (for our BLAGNs sample, the 
scatter is $0.4$). Note that, in this step, we added $0.3$ dex uncertainty 
to $\dot{M}$ and $0.17$ dex uncertainty to SFR to mimic approximately 
the measurement error in each variable (see Section~\ref{sec:err}). We 
repeated the selection $10^6$ times and constructed the distribution 
of $\rho$. We found that, of all the Eddington-ratio distributions tested, 
the $\sigma_L=0.4$ model comes closest to reproducing the observed 
anti-correlation between $\mathrm{s}\dot{M}/\mathrm{sSFR}$ and 
$M_{\rm BH}-M_{\ast}$, albeit with a probability of 
only $\sim 0.5\%$. The simulations also indicate that the probability to 
reproduce the observed scatter of the distribution of $\dot{M}/\mathrm{SFR}$ 
is only $\sim 0.05\%$. Note that in the $\sigma_{\rm L}=0.4$ model, the 
scatter between AGN luminosity and SFR is only $\sim 0.6$ dex. Broader 
Eddington-ratio distributions (e.g., uniform or shallow power-law) decrease 
the likelihood of selection effects reproducing a constant $\dot{M}$/SFR 
ratio, as do larger values of $\sigma_\mu$ or $\sigma_{\rm F}$. Moreover, 
we also performed the same simulations to the high- and low-redshift 
sub-samples and found the likelihood of selection effects also decrease. 

From the simulations, we can conclude that the observed anti-correlation 
between specific growth ratio and mass ratio and/or the narrow distribution 
of $\dot{M}/\rm{SFR}$ are unlikely to be solely due to selection 
effects but instead support the idea that there is a (instantaneous) 
connection between star formation and SMBH accretion activity (see the lower 
panels of Figure~\ref{fig:grms}). However, this conclusion rests upon the 
(rather uncertain) assumed Eddington-ratio distribution, as well as the 
similarly uncertain intrinsic scatter in $M_{\rm BH}$ and 
$M_{\rm BH}-M_{\ast}$. Better estimates of these quantities and their 
scatter are necessary to fully confirm that our results are not 
purely a selection effect. 

AGN feedback may explain the anti-correlation between 
$\mathrm{s}\dot{M}/\mathrm{sSFR}$ and $M_{\rm BH}/M_*$ and 
the constant ratio of $\dot{M}$/SFR. As has been long suggested, 
AGN feedback may regulate star formation and therefore maintain 
a correlation between AGN accretion and galaxy star formation 
\citep[e.g.,][]{sil98, kin03, dmt05, hop06}. However, recent 
simulations with a more realistic ISM structure suggest that AGN 
feedback can inject substantial energy into the ISM without having 
significant effect on cold gas in the galaxy 
\citep[e.g.,][]{bou14,gab14,roo14}. If so, our results are difficult to 
explain by AGN feedback. 

Our results can also be understood in a scenario where AGN activity 
and star formation are both governed by cold gas supply, without a 
need for AGN feedback. This scenario could lead to a connection 
between the AGN luminosity and SFR or even an AGN ``main sequence'' 
\citep[e.g.,][]{mul12, ros13a, ros13b, vit14}. For a host galaxy with an 
under-massive SMBH, we would expect the specific SMBH mass growth 
rate to be larger than that of the host galaxy since the mean ratio of the 
SFR to AGN luminosity remains roughly constant with changing $M_{\ast}$ 
\citep{mul12}. That is, an under-massive SMBH has a 
preferentially  steeper evolutionary vector than an over-massive 
SMBH on the $M_{\rm{BH}}-M_{\ast}$ plane.

\subsection{The AGN Duty Cycle among Star-Forming Galaxies}
\label{sec:dis3}
We are also interested in the AGN duty cycle and star formation 
timescale, $\tau_{\rm SF}$, that would maintain the $M_{\rm{BH}}-M_{\ast}$ 
relation as our sources evolve forward in time. We started with a 
Monte Carlo simulation which re-sampled $M_{\rm BH}$, $M_{\ast}$, 
$\dot{M}$, and $\rm{SFR}$ based on the observed values and errors 
for every source in our sample. For each realization, we evolved 
every source in time and calculated their new positions in the 
$M_{\rm{BH}}-M_{\ast}$ plane by integrating 
$\mathrm{SFR}(t)$ and $\dot{M}(t)$ over a grid of AGN duty cycles 
and $\tau_{\rm SF}$. For the high- (low-) redshift sub-sample, all 
sources were evolved from their current redshifts to $z=1$ ($z=0.2$). 
$\mathrm{SFR}(t)$ was assumed to be 
$\mathrm{SFR}(t)=\mathrm{SFR}(t_0)\exp(-(t-t_0)/\tau_{\rm{SF}})$, 
where $\mathrm{SFR}(t_0)$ was the re-sampled $\mathrm{SFR}$. As for 
$\dot{M}(t)$, we assumed that, on average (e.g., over $\sim 100\ 
\rm{Myr}$), $\dot{M}(t)=\dot{M}(t_0)\exp(-(t-t_0)/\tau_{\rm{AGN}})$. 
That is, following similar derivations in the previous section, 
the new positions of our sources are ($M_{\rm{BH,new}}$, 
$M_{\ast, \rm{new}}$), where $M_{\rm{BH,new}}=M_{\rm{BH,init}}+\int 
\dot{M}(t)dt\sim M_{\rm{BH,init}}+\tau_{\rm{AGN}}\dot{M}(t_0)$ 
and $M_{\ast,\rm{new}}=M_{\ast,\rm{init}}+\int 
\mathrm{SFR}(t)dt\sim M_{\ast,\rm{init}}+\tau_{\rm{SF}}\mathrm{SFR}(t_0)$ 
(as long as the evolution timescale is much larger than $\tau_{\rm SF}$). 
$M_{\rm{BH,init}}$ and $M_{\ast,\rm{init}}$ are the observed SMBH mass 
and galaxy stellar mass, respectively. We define the AGN duty cycle as 
$\tau_{\rm AGN}/\tau_{\rm SF}$. Assuming our AGN hosts are drawn from 
the star-forming main sequence ---as suggested by \cite{ros13b}--- this 
definition of AGN duty cycle essentially measures the fraction of BLAGNs 
among star-forming galaxies (we will discuss this definition later). We 
then determined the fraction ($f_{\rm on}$) of our sources whose new 
positions are consistent with the biased HR04 mass ratio (see 
Section~\ref{sec:bias}) and its $1\sigma$ uncertainty for each combination 
of AGN duty cycle and $\tau_{\rm SF}$. Finally, we found the AGN duty cycle 
and $\tau_{\rm SF}$ that result in the highest $f_{\rm on}$. After $10^4$ 
re-sampled realizations, we obtained the distribution of the AGN duty cycle 
and $\tau_{\rm SF}$ that would best maintain the $M_{\rm{BH}}-M_{\ast}$ 
relation. 

The duty cycle used here is only equivalent to 
$\mathrm{sSFR}/\mathrm{s}\dot{M}$ or $\mathrm{SFR}/\dot{M}$ if 
$\mathrm{s}\dot{M}$ and $\mathrm{sSFR}$ are both very small (i.e., 
if $\mathrm{s}\dot{M}\tau_{\rm AGN}$ and $\mathrm{sSFR}\tau_{\rm SF}$ 
both are $\ll 1$, then, $\Delta(\log M_{\rm{BH}}/M_{\ast})\approx \ln(10) 
(\mathrm{s}\dot{M}\tau_{\rm AGN} - \mathrm{sSFR}\tau_{\rm SF})$). This 
is true for our low-redshift sample. If, however, 
$\mathrm{s}\dot{M}\tau_{\rm AGN}$ and/or $\mathrm{sSFR}\tau_{\rm SF}$ 
are $\sim 1$ (e.g., our high-redshift sample), the AGN duty cycle is a 
complicated function of $M_{\rm BH}$, $M_{\ast}$, $\dot{M}$, and SFR. 
Actually, for our assumed SMBH accretion and star formation histories, 
our definition of the duty cycle measures the ratio of the 
lifetime of BLAGNs to that of active star formation. 

\begin{figure}
\plotone{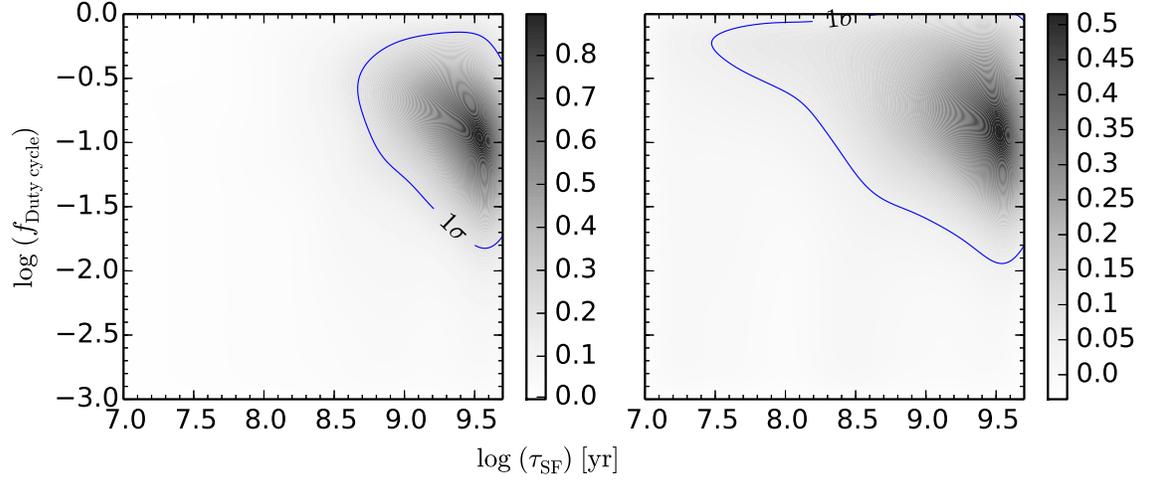}
\caption{
The PDF of $\tau_{\rm{SF}}$ and $\rm AGN\ duty\ cycle$ (which measures the 
fraction of BLAGNs among star-forming galaxies, see Section~\ref{sec:dis3}) 
that would result in the largest fraction of our sources lying on the local 
SMBH mass-bulge mass relation. Left (right) panels are for the high-(low-) 
redshift sub-samples. 
The contour in each panel indicates the $1\sigma$ (i.e., $68.3\%$) 
uncertainty. The color map in each panel indicates the probability 
density. 
}
\label{fig:duty}
\end{figure}

The two-dimensional PDF of the AGN duty cycle and $\tau_{\rm SF}$ is 
plotted in Figure~\ref{fig:duty} for the high-redshift (left panel) 
and low-redshift (right panel) sub-samples. The blue contour in each 
panel represents the``$1\sigma$'' scatter of the most likely AGN duty 
cycle and $\tau_{\rm SF}$. That is, given the measurement errors of 
$M_{\rm BH}$, $M_{\ast}$, $\dot{M}$, and $\rm{SFR}$, there is a $68.3\%$ 
probability that AGN duty cycles and $\tau_{\rm SF}$ which best maintain 
the $M_{\rm{BH}}-M_{\ast}$ relation are within the contour. 
To estimate the one-dimensional distribution of the AGN duty cycle 
(regardless of $\tau_{\rm SF}$), we integrated the two-dimensional 
PDF along $\tau_{\rm SF}$. From this distribution, the preferred 
AGN duty cycle is $\sim0.1$  \citep[agreeing with][]{sil09}
in both the high- and low-redshift sub-samples ($0.1_{-0.08}^{+0.24}$ 
at $z>1$, and $0.1_{-0.09}^{+0.32}$ at $z<1$, with errors defined 
by the 15.87 and 84.13 percentiles). In other words, our data favor 
a non-evolving (by no more than a factor of $\sim 4$) AGN duty cycle 
of $\sim 10\%$ for star-forming galaxies at both $z\sim 1.5$ and 
$z\sim 0.5$, ruling out very high ($\sim 67.6\%$) duty cycles at the 
$95 \%$ confidence level. The large confidence intervals for the AGN 
duty cycle are partly due to the large uncertainties in the 
estimated masses and growth rates. There may also be a large intrinsic 
scatter in AGN duty cycle between different galaxies. We note 
that these duty cycles are appropriate only for rapidly accreting 
SMBHs (i.e., BLAGNs or other AGNs with similar Eddington ratios) in 
star-forming galaxies.  Lower-luminosity AGNs are likely to have higher 
duty cycles, while quiescent galaxies \citep[known to have lower AGN 
fractions, e.g.,][]{ros13a, ros13b, tru13} are likely to have lower 
duty cycles \citep[for example, as found by][]{bon12}.

\section{Summary and Future Work}
\label{sec:summ}
\subsection{Summary}
In this paper, we studied 69 BLAGNs selected from the COSMOS and 
CDF-S fields based on X-ray observations, FIR observations, high-quality 
spectroscopy, and multi-band photometry. With these data, we 
simultaneously determined $M_{\rm BH}$, $M_{\ast}$, $\dot{M}$, 
and SFR and investigated the evolution of the $M_{\rm{BH}}-M_{\ast}$ 
relation. Our main conclusions are the following:

(1) Our sample suggests that, up to $z\sim 2.0$, there is no evolution  
(no more than $\sim 0.2$ dex) of the $M_{\rm BH}-M_{\ast}$ 
relation. The $M_{\rm BH}-M_{\rm Bul}$ relation, however, is likely 
to evolve, simply because our galaxies are expected to be more 
disk-dominated than the local sample of HR04. See Sections~\ref{sec:msevol} 
\&~\ref{sec:dis1}. 

(2) Our data indicate an anti-correlation between $M_{\rm BH}/M_{\ast}$ 
and the ratio of the two specific mass growth rates. Such an anti-correlation 
suggests that the on-going AGN activity and host-galaxy star formation 
are physically connected. Under appropriate assumptions about the 
AGN accretion and galaxy star formation histories, this anti-correlation 
can maintain the $M_{\rm BH}/M_{\ast}$ relation. See 
Sections~\ref{sec:tmevol} \&~\ref{sec:dis2}.

(3) We also investigate the possible values of AGN duty cycle which best 
maintain the non-evolving $M_{\rm BH}-M_{\ast}$ relation. Our data favor 
a non-evolving (i.e., a factor of $\lesssim 4$) AGN duty 
cycle of about $0.1$ for rapidly accreting SMBHs in star-forming galaxies 
($0.1_{-0.08}^{+0.24}$ at $z>1$, and $0.1_{-0.09}^{+0.32}$ at $z<1$). 
See Section~\ref{sec:dis3}. 

These results fit into a picture where the same gas reservoir fuels both 
AGN activity and galactic star formation. 

\subsection{Future Work}
Our results can be advanced in several regards. First, our sample size 
is significantly limited by the \textit{Herschel} sensitivity. Only 
$\sim 1/3$ of BLAGNs are detected by \textit{Herschel}. In this work 
we attempt to correct for this bias, but it would be better to simply 
have a less-biased sample. Future deeper rest-frame FIR 
observations (e.g., with ALMA, SCUBA-2, SPICA) can enlarge the sample 
size and are crucial for reducing the uncertainty and improving our 
understanding of the connection between AGN activity and star formation. 
As we have pointed out in Section~\ref{sec:bias}, the biases we mentioned 
(e.g., the bias of $M_{\rm{BH}}$ alone and the bias of 
$M_{\rm{BH}}/M_{\ast}$) depend on this connection. Therefore, deeper 
FIR surveys can also help us better address the selection biases in this 
work. This improvement, in turn, providing benefits to both our 
understanding of the co-evolution of SMBHs and their host galaxies (e.g., 
the formation of the local $M_{\rm{BH}}-M_{\rm{Bul}}$ relation) and SMBH 
mass demographics. 

Also, the large intrinsic scatter of $M_{\rm{BH}}$ prevents 
us from obtaining higher confidence level conclusions. Future 
improvements of the single-epoch virial estimators will be 
very helpful \citep[for example, with large multi-object 
reverberation mapping campaigns][]{she14}. 

Finally, more detailed observations and studies of physical properties 
(other than $M_{\ast}$) of the host galaxies are also very important. 
These properties (e.g., morphology and gas content) can help place 
additional distinct constraints on the co-evolution path of SMBHs and 
their host galaxies (e.g, the bulge-to-total ratio and the AGN/star 
formation triggering mechanism).

\acknowledgments 
We would like to thank the anonymous referee for his/her helpful feedback 
that improved this work. 
We thank Angela Bolzonella, Michael Eracleous, He Gao, Kazumi Kashiyama, 
Yuexing Li,  Jessie Runnoe, Nicholas Ross, Yue Shen, Sydney Sherman and 
Guang Yang for beneficial discussions. 

MYS acknowledges support from the China Scholarship Council (No.\ [2013]3009), 
and the National Natural Science Foundation of China under grant \#11222328. 
JRT acknowledges support by NASA through Hubble Fellowship grant \#51330 
awarded by the Space Telescope Science Institute. WNB and BL acknowledge 
support from \textit{Chandra} X-ray Center grant AR3-14015 and NASA ADP 
grant NNX10AC99G. DMA gratefully acknowledges support from the Science 
and Technology Facilities Council through grant ST/I001573/1 and the Leverhulme 
Trust. YQX acknowledges support of the Thousand Young Talents program 
(KJ2030220004), the 973 Program (2015CB857004), the USTC startup funding 
(ZC9850290195), the National Natural Science Foundation of China (NSFC-11473026, 
11421303), and the Strategic Priority Research Program ``The Emergence of 
Cosmological Structures'' of the Chinese Academy of Sciences (XDB09000000).

This research has made use of data from the public data release of the 
PACS Evolutionary Probe PEP \citep{lut11} and the GOODS--\textit{Herschel} 
programs \citep{elb11} as described in \cite{mag13}. 
This research has also made use of data from the HerMES project 
(http://hermes.sussex.ac.uk/). HerMES is a \textit{Herschel} Key 
Programme utilising Guaranteed Time from the SPIRE instrument team, 
ESAC scientists and a mission scientist. The HerMES data were accessed 
through the \textit{Herschel} Database in Marseille (HeDaM - http://hedam.lam.fr) 
operated by CeSAM and hosted by the Laboratoire d'Astrophysique de Marseille. 

We gratefully acknowledge the contributions of the entire COSMOS collaboration 
consisting of more than 80 scientists. More information on the COSMOS survey is 
available at http://www.astro.caltech.edu/cosmos. This work has made 
use of the NASA/IPAC Infrared Science Archive, which is operated by the 
Jet Propulsion Laboratory, California Institute of Technology, under 
contract with the National Aeronautics and Space Administration. 

This work has made use of catalogs from the 3D-HST Treasury Program 
(GO 12177 and 12328) using observations from the NASA/ESA HST, which 
is operated by the Association of Universities for Research in Astronomy, 
Inc., under NASA contract NAS5-26555. This work has made use of 
Astropy, a community-developed core Python package for Astronomy 
(Astropy Collaboration, 2013).

\clearpage

\clearpage
\begin{deluxetable}{ccc}
\tabletypesize{\scriptsize}
\tablecolumns{3}
\tablecaption{Summary of Sample Selection \label{tab1}}
\tablewidth{0pt}
\tablehead{
\colhead{Selection Criteria} & \colhead{COSMOS} & \colhead{CDF-S}}
\startdata
Total $z<2.4$ X-ray BLAGNs  &  221  &  30 \\
X-ray BLAGNs with optical/UV and \textit{Herschel} photometry  &  85  &  11 \\
Final sample: X-ray BLAGNs with optical/UV photometry, robust IR luminosity, 
good spectra  &  62  &  7 \\
\enddata
\end{deluxetable}

\clearpage
\LongTables
\begin{deluxetable}{ccccccccccc}
\tabletypesize{\scriptsize}
\tablecolumns{11}
\tablecaption{Summary of Properties of AGNs and Host Galaxies \label{tab2}}
\tablewidth{0pt}
\tablehead{
\colhead{ID} & \colhead{RA} & \colhead{DEC} & \colhead{$z$} & 
\colhead{SF\tablenotemark{a}} & \colhead{Field} & \colhead{$\log M_{\rm{BH}}$} &
\colhead{$\log M_{\ast}$} & \colhead{$\log L_{\rm{Bol}}$\tablenotemark{b}} &
\colhead{$\log L_{\rm{IR}}$\tablenotemark{d}} &
\colhead{$\Delta(\log L_{\rm{IR}})$\tablenotemark{e}}\\
\colhead{} & \colhead{deg} & \colhead{deg} & \colhead{} & 
\colhead{} & \colhead{} & \colhead{$M_{\odot}$} &
\colhead{$M_{\odot}$} & \colhead{$\rm{erg\ s^{-1}}$} &
\colhead{$\rm{erg\ s^{-1}}$} &
\colhead{$\rm{erg\ s^{-1}}$}} 
\startdata
1  &  149.83875  &  2.67511  &  0.26  &  S  &  COSMOS  &  7.07  &  10.88  &  44.51  &  44.93  &  0.06  \\ 
2  &  150.15837  &  2.13961  &  1.83  &  Z  &  COSMOS  &  8.99  &  11.75  &  46.08  &  46.74  &  0.07  \\
3  &  150.34592  &  2.14758  &  1.26  &  Z  &  COSMOS  &  8.28   &  11.23   &  45.38\tablenotemark{c}   &  46.17   &  0.08 \\
4  &  150.0425  &  2.62917  &  1.57  &  Z  &  COSMOS  &  8.68   &  11.90   &  46.16   &  46.44   &  0.05  \\
5  &  149.69879  &  2.44122  &  1.52  &  I  &  COSMOS  &  8.74   &  10.52   &  45.77   &  46.17   &  0.07  \\
6  &  149.881  &  2.45083  &  1.32  &  Z  &  COSMOS  &  8.01   &  11.05   &  45.28   &  46.11   &  0.02  \\
7  &  149.99158  &  1.72428  &  1.62  &  S  &  COSMOS  &  8.19   &  11.55   &  46.02   &  46.23   &  0.04  \\
8  &  150.30954  &  2.39914  &  1.80  &  S  &  COSMOS  &  8.81   &  11.49   &  46.46   &  46.35   &  0.04  \\
9  &  150.62525  &  1.80289  &  0.63  &  I  &  COSMOS  &  7.43   &  10.79   &  45.90   &  45.67   &  0.09  \\
10  &  150.05871  &  2.47739  &  1.26  &  Z  &  COSMOS  &  8.58   &  11.52   &  45.66   &  46.03   &  0.03  \\
11  &  149.70587  &  2.41975  &  1.12  &  Z  &  COSMOS  &  9.36   &  11.37   &  45.77   &  45.92   &  0.03  \\
12  &  149.94171  &  2.79544  &  1.07  &  S  &  COSMOS  &  8.67   &  11.38   &  46.19   &  45.90   &  0.07  \\
13  &  149.58283  &  2.48433  &  0.34  &  S  &  COSMOS  &  8.15   &  11.25   &  45.01   &  44.76   &  0.01  \\
14  &  150.12508  &  2.86175  &  1.58  &  I  &  COSMOS  &  8.30   &  11.80   &  46.08\tablenotemark{c}   &  45.90   &  0.06  \\
15  &  150.32737  &  2.46094  &  1.05  &  I  &  COSMOS  &  8.70   &  11.29   &  45.73   &  45.75   &  0.10  \\
16  &  149.59246  &  1.75675  &  1.96  &  S  &  COSMOS  &  8.90   &  11.23   &  47.15   &  46.66   &  0.26  \\
17  &  150.14708  &  2.71747  &  1.18  &  Z  &  COSMOS  &  8.13   &  10.79   &  45.05   &  45.63   &  0.12  \\
18  &  150.19558  &  2.00442  &  1.92  &  Z  &  COSMOS  &  9.02   &  11.14   &  46.20   &  46.22   &  0.11  \\
19  &  150.63387  &  2.59369  &  0.66  &  S  &  COSMOS  &  8.01   &  10.12   &  45.59   &  45.42   &  0.06  \\
20  &  150.49567  &  2.41256  &  1.37  &  Z  &  COSMOS  &  8.43   &  11.19   &  45.37   &  45.64   &  0.16  \\
21  &  150.57633  &  2.18142  &  0.55  &  I  &  COSMOS  &  8.29   &  11.06   &  44.48   &  44.93   &  0.00  \\
22  &  150.64521  &  2.71481  &  0.20  &  S  &  COSMOS  &  6.55   &  10.63   &  43.81   &  44.07   &  0.04  \\
23  &  150.19504  &  1.79383  &  1.87  &  I  &  COSMOS  &  8.74   &  11.57   &  45.69   &  46.09   &  0.10  \\
24  &  149.41992  &  2.03553  &  1.48  &  I  &  COSMOS  &  9.12   &  11.49   &  45.74   &  45.84   &  0.10  \\
25  &  150.58187  &  2.28769  &  1.34  &  Z  &  COSMOS  &  7.99   &  10.83   &  45.72   &  45.90   &  0.09  \\
26  &  149.47792  &  2.64247  &  1.6  &  S  &  COSMOS  &  8.64   &  11.77   &  46.26   &  45.94   &  0.22  \\
27  &  149.86558  &  2.00306  &  1.25  &  Z  &  COSMOS  &  8.11   &  10.41   &  45.19   &  45.47   &  0.10  \\
28  &  149.64183  &  2.74089  &  1.89  &  S  &  COSMOS  &  8.72   &  10.18   &  46.45   &  46.03   &  0.24  \\
29  &  150.658  &  2.7835  &  0.21  &  I  &  COSMOS  &  6.95   &  10.66   &  44.85   &  44.29   &  0.04  \\
30  &  150.55487  &  2.641  &  1.14  &  Z  &  COSMOS  &  8.51   &  10.76   &  45.33   &  45.14   &  0.35  \\
31  &  149.58521  &  2.05111  &  1.35  &  Z  &  COSMOS  &  8.80   &  11.28   &  45.93   &  45.77   &  0.29  \\
32  &  150.19975  &  2.19089  &  1.51  &  Z  &  COSMOS  &  8.89   &  11.18   &  45.56   &  45.57   &  0.16  \\
33  &  150.11929  &  2.85353  &  0.77  &  I  &  COSMOS  &  9.09   &  11.12   &  45.60   &  45.12   &  0.07  \\
34  &  150.31367  &  2.80361  &  1.46  &  I  &  COSMOS  &  9.06   &  10.80   &  45.69   &  45.58   &  0.25  \\
35  &  149.91692  &  2.38522  &  1.13  &  I  &  COSMOS  &  8.59   &  11.42   &  45.68   &  45.41   &  0.22  \\
36  &  149.78971  &  2.32125  &  0.38  &  I  &  COSMOS  &  7.39   &  10.55   &  44.63   &  44.14   &  0.01  \\
37  &  150.6355  &  1.66917  &  1.79  &  I  &  COSMOS  &  8.78   &  10.33   &  45.74   &  45.31   &  0.62  \\
38  &  150.30004  &  2.70936  &  0.73  &  I  &  COSMOS  &  7.61   &  9.33   &  44.50\tablenotemark{c}   &  44.77   &  0.09  \\
39  &  150.13954  &  1.6365  &  0.52  &  Z  &  COSMOS  &  7.49   &  10.30   &  44.53   &  43.93   &  1.41  \\
40  &  150.23083  &  2.57817  &  1.40  &  Z  &  COSMOS  &  8.87   &  11.35   &  45.87   &  46.02   &  0.20  \\
41  &  150.80187  &  2.00061  &  1.78  &  I  &  COSMOS  &  9.21   &  10.50   &  45.47\tablenotemark{c}   &  46.06   &  0.07  \\
42  &  150.62771  &  2.741  &  0.82  &  I  &  COSMOS  &  7.92   &  10.50   &  45.43   &  45.12   &  0.10  \\
43  &  150.384  &  1.57247  &  1.36  &  S  &  COSMOS  &  8.17   &  11.09   &  45.92   &  45.85   &  0.11  \\
44  &  150.25267  &  2.48642  &  2.04  &  I  &  COSMOS  &  9.06   &  11.39   &  45.89   &  46.16   &  0.13  \\
45  &  150.24517  &  1.90008  &  1.56  &  S  &  COSMOS  &  8.66   &  11.18   &  45.97   &  45.97   &  0.04  \\
46  &  150.19467  &  2.06792  &  0.55  &  I  &  COSMOS  &  7.29   &  11.17   &  44.93   &  44.77   &  0.08  \\
47  &  150.13908  &  1.877  &  0.83  &  I  &  COSMOS  &  8.12   &  11.07   &  45.18   &  45.09   &  0.01  \\
48  &  150.006  &  2.81242  &  0.77  &  S  &  COSMOS  &  8.38   &  10.77   &  45.59   &  45.10   &  0.10  \\
49  &  149.86796  &  2.35192  &  0.35  &  I  &  COSMOS  &  6.84   &  10.63   &  44.09   &  44.51   &  0.02  \\
50  &  149.62429  &  2.18067  &  1.19  &  I  &  COSMOS  &  8.43   &  10.99   &  45.55   &  45.57   &  0.13  \\
51  &  149.4795  &  2.80183  &  1.11  &  S  &  COSMOS  &  8.68   &  11.29   &  45.97   &  45.37   &  0.19  \\
52  &  149.868  &  2.33075  &  1.49  &  I  &  COSMOS  &  8.67   &  11.18   &  45.67   &  46.10   &  0.07  \\
53  &  150.55046  &  1.709  &  0.37  &  I  &  COSMOS  &  7.35   &  10.44   &  44.49   &  44.34   &  0.08  \\
54  &  149.6692  &  2.07406  &  0.34  &  Z  &  COSMOS  &  8.31   &  10.67   &  44.42   &  44.17   &  0.10  \\
55  &  150.10521  &  1.98117  &  0.37  &  S  &  COSMOS  &  8.12   &  8.75   &  45.23   &  44.18   &  0.13  \\
56  &  150.05379  &  2.58967  &  0.7  &  S  &  COSMOS  &  8.33   &  11.21   &  45.78   &  44.88   &  0.11  \\
57  &  150.1237  &  2.35825  &  0.73  &  Z  &  COSMOS  &  7.21   &  10.62   &  44.83   &  44.91   &  0.11  \\
58  &  150.68317  &  2.57461  &  0.38  &  Z  &  COSMOS  &  7.75   &  10.31   &  44.78   &  44.07   &  0.12  \\
%59  &  150.42192  &  2.38556  &  1.51  &  I  &  COSMOS  &  8.17   &  10.89   &  45.03   &  45.68   &  0.12  \\
59  &  150.2082  &  1.87536  &  1.16  &  I  &  COSMOS  &  8.83   &  10.74   &  45.36   &  45.31   &  0.14  \\
60  &  149.89484  &  2.17444  &  1.32  &  Z  &  COSMOS  &  8.57   &  10.90   &  45.19   &  45.75   &  0.09  \\
61  &  149.76346  &  2.33411  &  1.13  &  Z  &  COSMOS  &  8.08   &  10.94   &  45.28   &  45.37   &  0.13  \\
62  &  149.66362  &  2.08519  &  1.22  &  Z  &  COSMOS  &  8.19   &  10.80   &  45.54   &  45.48   &  0.13  \\
63  &  53.22033  &  -27.85556  &  1.23  &  S04  &  CDF-S  &  8.25   &  10.74   &  45.06   &  44.96   &  0.07  \\
64  &  53.11037  &  -27.67658  &  1.03  &  S04  &  CDF-S  &  7.80   &  10.64   &  45.77   &  45.55   &  0.02  \\
65  &  53.10483  &  -27.70525  &  1.62  &  S04  &  CDF-S  &  8.74   &  11.00   &  45.64   &  46.15   &  0.04  \\
66  &  53.07146  &  -27.71761  &  0.57  &  S04  &  CDF-S  &  7.65   &  10.35   &  44.72   &  45.33   &  0.14  \\
67  &  53.0455  &  -27.73756  &  1.615  &  S04  &  CDF-S  &  8.84   &  11.06   &  45.58   &  45.58   &  0.01  \\
68  &  53.25642  &  -27.76183  &  0.62  &  S04  &  CDF-S  &  7.96   &  10.82   &  44.72   &  44.85   &  0.03  \\
69  &  53.12525  &  -27.75656  &  0.96  &  S04  &  CDF-S  &  7.43   &  10.67   &  44.75   &  45.59   &  0.02  \\

\enddata
\tablecomments{}
\tablenotetext{a}{Spectrum source flag: ``S'', ``I'' and ``Z'' indicate the spectrum and redshift are 
from the SDSS archive, the COSMOS Magellan/IMACS campaign \citep{tru09a} and the zCOSMOS VLT/VIMOS 
campaign \citep{lil07}, respectively; ``S04'' means the spectrum and redshift are from \cite{szo04}.}
\tablenotetext{b}{$\dot{M}$ can be calculated from $L_{\rm{Bol}}$ using Equation~\ref{eq:mdot}.}
\tablenotetext{c}{Objects that are detected in the soft X-ray band but not in the hard X-ray band.}
\tablenotetext{d}{SFR can be calculated from $L_{\rm{IR}}$ using Equation~\ref{eq:msg}.}
\tablenotetext{e}{The uncertainty of the IR luminosity reported here is only relevant to the uncertainty 
of the FIR SED calibration. To fully account for the uncertainty of the IR luminosity, one should 
also add the intrinsic scatter of the SED template in quadrature (see Section~\ref{sec:err}).}
\end{deluxetable}

\end{document}